\documentclass{article}
\usepackage{jcappub} 
\usepackage{lineno}
\usepackage{journals}
\usepackage{orcidlink}

\arxivnumber{2403.04122} 

\usepackage{lipsum}
\usepackage{amsmath}
\usepackage{mathrsfs}
\usepackage[thinc]{esdiff}
\usepackage{physics}
\usepackage{booktabs}
\usepackage{threeparttable}
\usepackage{csvsimple-l3}

\usepackage{datatool}
\DTLsetseparator{,}

\DTLloaddb[noheader, keys={thekey,thevalue}]{data}{data.txt}
\newcommand{\data}[1]{\protect\DTLfetch{data}{thekey}{#1}{thevalue}}

\newcommand{\denunit}{\mathrm{GeV}\,\mathrm{cm^{-3}}}
\newcommand{\velunit}{\mathrm{km\,s^{-1}}}
\newcommand{\staudtcolor}{purple}
\newcommand{\maxwellcolor}{red}
\newcommand{\distribbandcolor}{grey}
\newcommand{\vcvarbandcolor}{purple}

\title{\LARGE{Sliding into DM: Determining the local dark matter density and speed distribution using only the local circular speed of the Galaxy}
}

\author[a]{Patrick G. Staudt$^{\orcidlink{0000-0002-5887-9738}}$,}
\author[a]{James S. Bullock$^{\orcidlink{0000-0003-4298-5082}}$,}
\author[b]{Michael Boylan-Kolchin$^{\orcidlink{0000-0002-9604-343X}}$,}
\author[a]{David Kirkby$^{\orcidlink{0000-0002-8828-5463}}$,}
\author[c]{Andrew Wetzel$^{\orcidlink{0000-0003-0603-8942}}$,}
\author[d]{and Xiaowei Ou$^{\orcidlink{0000-0002-4669-9967}}$}

\affiliation[a]{Department of Physics \& Astronomy, University of California, Irvine, California 92697, USA}
\affiliation[b]{Department of Astronomy, The University of Texas at Austin, Austin, TX 78712, USA}
\affiliation[c]{Department of Physics \& Astronomy, University of California, Davis, Davis, CA 95616, USA}
\affiliation[d]{Physics Department and Kavli Institute for Astrophysics and Space Research, Massachusetts Institute of Technology, 77 Massachusetts Avenue, Cambridge, MA 02139, USA}

\emailAdd{patrickstaudt1@gmail.com}

\abstract{
We use FIRE-2 zoom simulations of Milky Way size disk galaxies to derive easy-to-use relationships between the observed circular speed of the Galaxy at the Solar location, $v_\mathrm{c}$, and dark matter properties of relevance for direct detection experiments: the dark matter density, the dark matter velocity dispersion, and the speed distribution of dark matter particles near the Solar location. We find that both the local dark matter density and 3D velocity dispersion follow tight power laws with $v_\mathrm{c}$. Using this relation together with the observed circular speed of the Milky Way at the Solar radius, we infer the local dark matter density and velocity dispersion near the Sun to be $\rho = \data{rho_GeV}\pm\data{drho_GeV}\ \denunit$ and $\sigma_{\rm 3D} = \data{disp}^{+\data{ddisp_plus}}_{-\data{ddisp_minus}}\,\mathrm{km\,s^{-1}}$. We also find that the distribution of dark matter particle speeds is well-described by a modified Maxwellian with two shape parameters, both of which correlate with the observed $v_{\rm c}$. We use that modified Maxwellian to predict the speed distribution of dark matter near the Sun and find that it peaks at a most probable speed of $\data{v0_mw}\,\mathrm{km\,s^{-1}}$ and begins to truncate sharply above $\data{vcrit_mw}\,\mathrm{km\,s^{-1}}$. This peak speed is somewhat higher than expected from the standard halo model, and the truncation occurs well below the formal escape speed to infinity, with fewer very-high-speed particles than assumed in the standard halo model.
}


\begin{document}
\begin{filecontents*}{data.txt}
maxdiff,11.1\%
logrho,7.04
dlogrho,0.06
dlogrho_plus,n/a
dlogrho_minus,n/a
rho_GeV,0.42
drho_GeV,0.06
drho_GeV_plus,n/a
drho_GeV_minus,n/a
disp,280
ddisp,n/a
rho_canon_GeV,0.39
drho_canon_GeV,0.09
logrho_canon,7.01
dlogrho_canon_plus,0.09
dlogrho_canon_minus,0.11
dlogrho_canon,n/a
ddisp_plus,19
ddisp_minus,18
disp_slope,0.81
ddisp_slope,0.09
ddisp_slope_plus,n/a
ddisp_slope_minus,n/a
disp_amp,143
ddisp_amp,n/a
ddisp_amp_plus,11
ddisp_amp_minus,10
den_slope,0.80
dden_slope,0.19
dden_slope_plus,n/a
dden_slope_minus,n/a
den_amp,5.2
dden_amp,0.5
dden_amp_plus,n/a
dden_amp_minus,n/a
logden_intercept,6.76
dlogden_intercept,0.07
dlogden_intercept_plus,n/a
dlogden_intercept_minus,n/a
d,111.0
dd,0.6
dd_plus,n/a
dd_minus,n/a
e,1.014
de,0.007
de_plus,n/a
de_minus,n/a
h,279.7
dh,1.9
dh_plus,n/a
dh_minus,n/a
j,0.588
dj,0.007
dj_plus,n/a
dj_minus,n/a
rho_1e7msun,1.11
drho_1e7msun,n/a
drho_1e7msun_plus,0.17
drho_1e7msun_minus,0.15
rho0_1e7msun,0.57
drho0_1e7msun,n/a
drho0_1e7msun_plus,0.10
drho0_1e7msun_minus,0.09
rho0_GeV,0.22
drho0_GeV,n/a
drho0_GeV_plus,0.04
drho0_GeV_minus,0.03
v0_mw,257
dv0_mw,8
dv0_mw_plus,n/a
dv0_mw_minus,n/a
vdamp_mw,455
dvdamp_mw,8
dvdamp_mw_plus,n/a
dvdamp_mw_minus,n/a
k,0.0350
dk,0.0004
dk_plus,n/a
dk_minus,n/a
vcrit_mw,470
maxdendiff,11.1\%
maxdispdiff,5.3\%
d_mao_lim_fit,40.6
dd_mao_lim_fit,n/a
dd_mao_lim_fit_plus,0.7
dd_mao_lim_fit_minus,0.8
p_mao_lim_fit,2.633
dp_mao_lim_fit,0.011
dp_mao_lim_fit_plus,n/a
dp_mao_lim_fit_minus,n/a
e_mao_lim_fit,3.10
de_mao_lim_fit,0.04
de_mao_lim_fit_plus,n/a
de_mao_lim_fit_minus,n/a
p_mao_naive_agg,1.961
stdev_linear_dendiff,4.4\%
stdev_linear_dispdiff,1.3\%
vesc_mw(vc),550
dvesc_mw(vc),30
dvesc_mw(vc)_plus,n/a
dvesc_mw(vc)_minus,n/a
veschat_amp,350
dveschat_amp,n/a
dveschat_amp_plus,30
dveschat_amp_minus,20
veschat_slope,0.54
dveschat_slope,0.08
dveschat_slope_plus,n/a
dveschat_slope_minus,n/a
vesc_min_std,1\%
vesc_max_std,4\%
mao_naive_rms,0.34\times10^{-3}
staudt_rms,0.26\times10^{-3}
mao_ours_rms,0.28\times10^{-3}
dp_mao_naive_agg,0.004
dp_mao_naive_agg_plus,n/a
dp_mao_naive_agg_minus,n/a
\end{filecontents*}

\maketitle
\flushbottom

\section{Introduction}
Dark matter accounts for more than eighty percent of the matter in the universe \citep{planck-collaboration2020planck}, but has managed to evade efforts to determine its precise makeup.
Among the most popular ideas is that dark matter is made up of as-of-yet undiscovered elementary particles of nature, and there are significant efforts underway to directly detect new particles with the characteristics necessary for them to be dark matter. 

 Weakly interacting massive particles, or WIMPs, are one of the leading candidates for particle dark matter. While so far attempts to directly detect WIMPs have been unsuccessful, the lack of signal provides important constraints on the allowed microphysical properties of theoretical dark matter particles.  In particular, direct detection experiments place joint limits on the allowed mass and nucleon-interaction cross section for particle dark matter (e.g. \citep{aprile2018dark,meng2021dark, aalbers2022first, aprile2023first}). Two astrophysical assumptions are critical for deriving these constraints (see ref. \cite{lewin1996review} for a nice discussion): 1) the local mass density of dark matter in the Earth's vicinity, $\rho$, and 2) the local speed distribution of dark matter particles, $f(v)$. Moreover, the interpretation of any future detection signal from such an experiment will rely heavily on both $\rho$ and $f(v)$. These facts motivate considerable effort to determine local dark matter properties as accurately as possible. Furthermore, in indirect detection,  the local dark matter density is used to normalize models for the dark matter density in the center of the Galaxy, and thus is fundamental in determining whether a WIMP dark matter model for the Galactic Center extended gamma-ray excess is compatible with the lack of annihilation signal from dwarf galaxies (e.g. \citep{abazajian2016bright, keeley2018what}).

Methods for determining the local dark matter density fall into two broad categories: Local and Global.  Local measures rely on tracers in the Solar vicinity, while Global estimates rely on fitting a density model to measurements taken throughout the Milky Way. See ref. \cite{read2014the-local} for a review of the various methods. We summarize fifteen estimates of $\rho$ in Table \ref{tab:rho} and see that, even among measurements that adopt similar approaches, estimates for the local density vary more than can be explained by the quoted statistical error bars. This suggests that there are systematic errors at play. There is also a trend for Local methods to give slightly higher dark matter densities than Global methods, with averages of $\rho = 0.47 \pm 0.11\ \denunit$ for Local and $\rho = 0.39 \pm 0.10\ \denunit$ for Global\footnote{The error bars on these averages are based on the standard deviation of the $N$ measurements in each group, weighted by $\sqrt{1 + 1/N}$.}. However, this is not an exhaustive literature search; more work could be done to determine whether there is a true bias with a physical basis. Ref. \cite{sivertsson2022estimating} found that assuming the Milky Way's disk is axisymmetric and in a steady state, as most Local measures do, can impart ${\sim}30\%$ uncertainty. Ref. \cite{widmark2021weighing}, which takes into account the Galaxy's time-varying structure and phase-space spiral, produces the lowest density estimate of the Local measurements we highlight. It will be interesting to see whether further Local measures that allow for time-varying structure and axi-asymmetry produce similar results. We further note that ref. \cite{salucci2010the-dark} adopted an approach qualitatively similar to the method we present below, in that they rely on only the local equation of centrifugal equilibrium, not the full rotation curve.  They find a preferred density of $\rho \simeq 0.43 \, \denunit$, which is similar to the overall average of the studies summarized in Table \ref{tab:rho}:   $\rho = 0.41 \pm 0.10 \, \denunit$. 

\begin{table*}
\centering
\addtolength{\leftskip} {-2cm}
\addtolength{\rightskip}{-2cm}
\caption{Twelve of the most recent studies measuring the local dark matter density $\rho$, grouped by Global and Local methods. The margins of error for the averages are calculated as $s\sqrt{1+1/N}$ where $s$ is the standard deviation of the measurements, and $N$ is the number of measurements. In deSales+19 \cite{de-salas2019on-the-estimation}, B1 and B2 refer to two different baryonic mass models.}
\label{tab:rho}
\begin{tabular}{llll}
\multicolumn{4}{c}{Global methods}\\
\cmidrule(lr){1-4}
                                                   & Model basis                                             & Note    & $\rho\,/\,[\denunit]$ \\
\cmidrule(lr){1-4}
Eilers+19 \citep{eilers2019the-circular}           & Fitting the circular speed curve                     &         & $0.30\pm0.03$\\
deSales+19 \citep{de-salas2019on-the-estimation}   & Fitting the circular speed curve                     & B1      & $0.30\pm0.03$\\
                                                   &                                                         & B2      & $0.38\pm0.04$\\
Nitschai+20 \citep{nitschai2020first}              & Anisotropic jeans modeling of disk stars                &         & $0.437\pm0.076$\\
Petac20 \citep{petac2020equilibrium}               & Galactic mass model with circular speed as one input & NFW     & $0.357^{+0.020}_{-0.021}$\\
                                                   &                                                         & Burkert & $0.381^{+0.020}_{-0.022}$\\
Benito+21 \citep{benito2021uncertainties}          & Fitting the circular speed curve                     &         & $0.6\pm0.1$\\
Hattori+21 \citep{hattori2021action-based}         & Galactic mass model from halo stars                     &         & $0.342\pm0.007$\\
Ou+24 \citep{ou2024the-dark}                       & Fitting the circular speed curve                     &         & $0.447\pm0.004$\\
\cmidrule(lr){1-4}
Average                                            &                                                         &         & $0.39\pm0.10$\\
                                                   &                                                         &         &\\
              
\multicolumn{4}{c}{Local methods}\\
\cmidrule(lr){1-4}
                                                   & Model basis                                             & Note    & $\rho\,/\,[\denunit]$ \\
\cmidrule(lr){1-4}
Sivertsson+18 \citep{sivertsson2018the-local}      & Vertical kinematics                                     &         & $0.46^{+0.07}_{-0.09}$                                  \\
Buch+19 \citep{buch2019using}                      & Very local kinematics                                   & A stars & $0.61\pm0.38$\\
                                                   &                                                         & G stars & $0.42^{+0.38}_{-0.34}$\\
Guo+20 \citep{guo2020measuring}                    & Vertical kinematics                                     &         & $0.50^{+0.09}_{-0.08}$                                   \\
Salomon+20 \citep{salomon2020kinematics}           & Vertical kinematics                                     &         & $0.37\pm0.09$                                           \\
Widmark+21 \citep{widmark2021weighing}             & Time-varying structure of the local phase-space spiral  &         & $0.32\pm0.15$\\
\cmidrule(lr){1-4}
Average       &                                                         &         & $0.47\pm0.11$\\
\\
\toprule
Average of all above &                                                    &         & $0.41\pm0.10\,\denunit$\\ \\
\toprule
\textbf{This paper}&                                                    &         & $\boldsymbol{\data{rho_GeV}\pm\data{drho_GeV}\,\denunit}$\\[0.7ex]
\end{tabular}
\end{table*}

For the speed distribution, the standard has been to assume a Maxwellian (Maxwell-Boltzmann) distribution that peaks at the Milky Way's local circular speed and truncates at the escape speed (e.g. \citep{kamionkowski1998galactic}). 
This assumption is often called the ``standard halo model" or SHM, usually with a circular speed of $220\,\mathrm{km\,s^{-1}}$. There has been extensive research into whether the SHM appropriately represents the Milky Way or even whether a Maxwellian is the correct functional form to adopt. Investigators typically carry out this evaluation on specific simulated galaxies, chosen from a larger pool of candidates based on how closely they match certain properties of the Milky Way \citep{bozorgnia2017implications}.

While some studies using high-resolution galaxy formation simulations have found speed distributions consistent with Maxwellians (e.g. ref. \citep{nunez-castineyra2019dark}),
a Maxwellian usually fails to reproduce the shape of dark matter speed distributions as seen in modern simulations, usually over-predicting the number of high-velocity particles \citep{ling2010dark,pillepich2014the-distribution,sloane2016assessing,kelso2016the-impact,butsky2016nihao, bozorgnia2016simulated}.
These studies tend to achieve better fits with more complex models, such as generalized Maxwellian distributions, Tsallis distributions, Eddington formulas, or various other modifications to the Maxwellian distribution such as those developed by ref. \cite{mao2013halo-to-halo} or \cite{lisanti2011dark}. The prevailing methodology has been to then analyze data from dark matter direct detection experiments assuming that the speed distribution for a specific simulated Milky Way analogue represents the Milky Way itself.

In this paper we analyze twelve zoom-in simulations of Milky Way size disk galaxies from the FIRE-2 project and attempt to develop formulaic methods that allow us to infer local dark matter properties of the Milky Way itself. We obviate the need to choose one particular simulated galaxy as a Milky Way analogue. Instead, we correlate the density, velocity dispersion, and speed distribution using a sliding scale of one single, simple observable: the local circular speed. 

In the next section we provide a short overview of dark matter detection and introduce some nomenclature.  In Section 3 we discuss our simulations, and in Section 4 we describe our analysis.  All of our results are presented in Section 5, and Section 6 is reserved for discussion and conclusions.

\section{Background and Nomenclature}
\label{sec:direct_detection}
The differential event rate for interactions between dark matter particles and a detector is
\begin{equation}
\label{eq:drde}
\dv{\mathcal{R}}{E_\mathrm{R}}=\frac{\rho}{m_\chi m_\mathrm{N}}\int_{v_\mathrm{min}}^\infty vf(v)\dv{\sigma_\mathrm{WN}(v,E_\mathrm{R})}{E_\mathrm{R}}\mathrm{d}^3v ,
\end{equation}
where $\rho$ is the local dark matter density, $m_\chi$ is the mass of the dark matter particle, $m_\mathrm{N}$ is the mass of the nucleus used in the detector, $\sigma_\mathrm{WN}$ is the interaction cross section between the nucleus and the dark matter particle, and $v_\mathrm{min}$ is the minimum dark matter particle speed required for the detector nucleus to recoil with energy $E_\mathrm{R}$. The value of $v_\mathrm{min}$  is given by 
\begin{equation}
\left(\frac{v_\mathrm{min}}{c}\right)^2={\frac{m_\mathrm{N}E_\mathrm{R}}{2\mu^2}}
.
\end{equation}
Here, $\mu$ is the reduced mass between the dark matter particle and the detector nucleon.
Note that for canonical leading order spin-independent DM-nucleus interactions with equal couplings of DM to protons and neutrons, the differential cross section is inversely proportional to the square of the WIMP's speed $\left(\mathrm{d}\sigma_\mathrm{WN}/\mathrm{d}E_\mathrm{R}\propto v^{-2}\right)$ \citep{marrodan-undagoitia2016dark, bozorgnia2017implications, evans2019refinement}. Therefore, the differential event rate's dependence on velocity is entirely captured by the halo integral
\begin{equation}
\label{eqn:halo_integral}
g(v_\mathrm{min}) = \int_{v_\mathrm{min}}^\infty\frac{f(v)}{v}\mathrm{d}^3v.
\end{equation}

A traditional assumption for the speed distribution of dark matter particles is a Maxwellian:
\begin{equation}
f(v)=\frac{1}{N(v_0)}\exp\left(-\frac{v^2}{v_0^2}\right),
\label{eqn:simple_max}
\end{equation}
where $v = |\vec{v}|$,  $v_0$ is the most-probable speed or ``peak speed", and
$ N(v_0) = \pi^{3/2}v_0^3$ is a normalization factor that gives $\int_0^\infty f(v)4\pi v^2dv=1$. It is common to  assume an isothermal sphere and set the peak speed equal to the circular speed of the Milky Way at the Solar radius: $v_0=v_\mathrm{c}$. Alternatively, $v_0$ can be expressed in terms of the 3D velocity dispersion:
\begin{equation}
v_0 = \sqrt{2/3}\,\sigma_\mathrm{3D}.
\label{eqn:23sigma}
\end{equation}
A slight modification of the Maxwellian assumption is to introduce a sharp truncation in Equation \ref{eqn:simple_max} above some escape speed, $v_\mathrm{esc}$.  In this case, the normalization factor must be modified:
\begin{equation}
N(v_0, v_{\rm esc}) = \pi v_0^2
\left[
\sqrt{\pi}v_0\erf
\left(\frac{v_\mathrm{esc}}{v_0}\right)
-2v_\mathrm{esc}\exp\left(-\frac{v_\mathrm{esc}^2}{v_0^2}\right)\right].
\end{equation}
An alternative to Equation \ref{eqn:simple_max} motivated by galaxy formation simulations is a parameterization presented by Mao et al. (2013) \citep{mao2013halo-to-halo}:
\begin{equation}
f(v) =
    \begin{cases}
        N\exp(-\frac{v}{v_0})(v^2_\mathrm{esc}-v^2)^p, & 0 \leq v\leq v_\mathrm{esc}\\
        0, & \text{otherwise}
    \end{cases}
\label{eqn:mao}
\end{equation}
where $p$ is a constant that is fit to simulation data. In Section \ref{sec:results}, we introduce a new model for $f(v)$ and tie its parameters to the circular speed. We then use the Milky Way's observed circular speed to determine its most likely parameters. In the Appendix Section \ref{sec:Mao}, we provide a comparison between our model and the functional form in Equation \ref{eqn:mao}.

\section{Simulations}
\label{sec:sims}
We analyze cosmological zoom-in simulations run with the \textsc{gizmo} code \citep{hopkins2015a-new-class} from the Feedback in Realistic Environments (FIRE) project\footnote{\url{https://fire.northwestern.edu}}. Specifically, we use the FIRE-2 model \citep{hopkins2018fire-2} to implement stellar feedback and star formation. An effective feedback implementation is important for regulating star formation, producing galaxies that are the correct mass for their dark matter halo size, and also in enabling the formation of galaxies with disk morphologies \citep{governato2004the-formation,guedes2011forming,hopkins2014galaxies, hopkins2018fire-2}. Additionally, feedback can affect galaxies' density profiles via ``feedback-induced core formation" (e.g. \cite{lazar2020a-dark} and references therein).  Galaxy formation is also known to increase the central dark matter velocity dispersion compared to dark-matter-only simulations \citep{mckeown2022amplified}, thus achieving realistic galaxy masses is crucial to producing dark matter properties that are accurate. The stellar feedback in FIRE includes stellar winds, radiation pressure, photoelectric heating, and photoheating from ionizing radiation. Each star particle assumes a Kroupa IMF \citep{kroupa2001on-the-variation}, has an age determined from its formation time, and inherits its metallicity from its parent gas particle. Within the gas particles, the simulations track 11 elemental abundances (H, He, C, N, O, Ne, Mg, Si, S, Ca, and Fe) with sub-grid diffusion via turbulence \citep{hopkins2016a-simple,su2017feedback,escala2018modelling}.  The conditions for star formation in gas are local self gravitation, sufficient density (${>}1000\rm\, cm^{-3}$), Jeans instability, and molecularity (following ref. \cite{krumholz2011a-comparison}).

 We focus on all galaxies in the FIRE-2 suite of ref. \cite{garrison-kimmel2019star}  with 1) stellar mass similar to that of the Milky Way and 2) a prominent disk component. This yields a sample of twelve, as described in more detail below. The initialization of these galaxies follows ref. \cite{onorbe2014how-to-zoom:} using the \textsc{music} package \citep{hahn2011multi-scale}. Six of the twelve come from the Latte suite \citep{wetzel2016reconciling,garrison-kimmel2017not-so-lumpy,hopkins2017anisotropic,garrison-kimmel2019the-local} and six from the ELVIS on FIRE project \citep{garrison-kimmel2019the-local,garrison-kimmel2019star}. The former is a collection of zooms on isolated galaxies each named with a unique letter and an ``m12" prefix, for example ``m12b". The latter consists of pairs with separations and relative velocities similar to the Milky Way-M31 pair.
 The FIRE collaboration generated these ELVIS pairs to investigate the effects of being in the Local Group. Given that Andromeda is itself Milky-Way like, both galaxies in each ELVIS pair are included in our analysis. Each ELVIS pair comprises a single zoom, and they follow the naming convention of famous duos. 

Dark matter particle mass is $3.5\times10^4\ \rm M_\odot$ in Latte and $\simeq2\times 10^4\ \rm M_\odot$ in ELVIS. Gas and star particles in Latte have initial masses of $7100\ \rm M_\odot$ while ELVIS simulations have approximately twice the resolution at 3500 to 4000 $\rm M_\odot$. Gas softening lengths are fully adaptive down to $\simeq0.5$--$1\,\rm pc$. The star-particle softening length is ${\simeq}4\,\rm pc$ physical. The dark matter force softening is ${\simeq}40\,\rm pc$ physical.

We initially considered all fourteen simulations in ref. \cite{garrison-kimmel2019star} with stellar masses similar to that the Milky Way ($10^{10-11}$ M$_{\odot}$).  Of these, we excluded two because they did not exhibit disks of the kind needed to measure the circular velocity: m12w and m12z.  We determined that these two galaxies were not disks using 1) visual appearance\footnote{\url{http://www.tapir.caltech.edu/~phopkins/Site/animations/a-gallery-of-milky-way-/}}, 2) a comparison of the ideal $\sqrt{GM/R_0}$ and the tangential component of cold gas velocity, and 3) thin-disk fraction as defined in ref. \cite{yu2021the-bursty}.
While m12z has a rotating component, its thin-disk fraction is only 5\%, and visually it does not resemble a disk. m12w, on the other hand, has a potentially acceptable thin-disk fraction of 23\%; however, visually it is an S0 lenticular and not a realistic Milky Way analogue. 

Note that in identifying the twelve galaxies we consider, we did not explicitly use the merger history, though large mergers would have likely resulted in non-disk systems and thus exclusion by our disk criterion.  The Milky Way itself appears to have had a fairly quiet merger history, with no major mergers within the last 10 Gyr. For example, the most significant merger event in the Milky Way's history (known as the `Kraken' merger), was a minor accretion $10.9^{+0.4}_{-0.7}$ Gyr ago (${\gtrsim}50$ dynamical times) with a halo mass ratio of only $1{:}7^{+4}_{-2}$ \citep{kruijssen2020kraken}. Dark matter components from this event have long since virialized, and any substructure created would have been washed out by the present day \citep{garrison-kimmel2017not-so-lumpy}.
Similarly, the Gaia-Enceladus (G-E) event occurred $9.1^{+0.4}_{-0.5}$ Gyr, or ${\sim}45$ dynamical times, ago. Ref. \cite{evans2019refinement} explored the effects of both including G-E in the Milky Way's speed distribution and updating $v_0$ and $v_\mathrm{esc}$ with more recent estimates. They did find an effect; however, the updated values of $v_0$ and $v_\mathrm{esc}$ were much more significant. 

With the above as context, the analysis we present below is merger-agnostic. In fact, our method intentionally departs from the ``single analogue method" of seeking a particular simulation that appears as close as possible to the Milky Way. Instead, we attempt to use a larger sample and make predictions based on a single observable, the local circular speed, without any selection other than having a disk to enable a circular-speed measurement. As with most correlations, the one we find does exhibit some noise.  The variation in merger history is one possible contributor to noise.  We note that a number of the FIRE-2 disks experienced past mergers similar to those inferred from the Milky Way \citep{panithanpaisal2021the-galaxy}, but dark matter substructure does not survive, as shown in ref. \cite{garrison-kimmel2017not-so-lumpy} and our Figure \ref{fig:disc_diffs} below. 
A future analysis, which would need to rely on a much larger simulation set, could potentially add observationally-oriented constraints to provide an even tighter prediction.

\section{Methods}
\label{sec:methods}

\subsection{Defining Sun-like Regions in the Simulations}
\label{sec:solar}
We are interested in exploring dark matter characteristics around mock Sun-like regions in the disks of our simulated galaxies.  In each simulation, the plane of the disk is defined as that perpendicular to the aggregate angular momentum of all gas, stars, and dark matter within 10 kpc of the galaxy's center. We find that the orientation of disk planes remains virtually identical if we include only stars when determining the angular momentum direction.

For analysis we use particles in ring-like regions centered on mock solar radii ($R_0 = 8.3$ kpc \citep{gillessen2009monitoring, mcmillan2011mass, schonrich2012galactic}) in each simulated disk. One could add a level of complexity to this study by scaling $R_0$ for each galaxy. However, the results below show that a fixed radius yields tight constraints on our conclusions, which we find satisfactory. We define each galaxy's solar ring by first taking a spherical shell 1.5 kpc thick with a midpoint radius 8.3 kpc from the center of the given galaxy. From there, we exclude the parts of the shell 0.5 kpc above and below the plane of the disk. These choices for thickness and height allow the ring to be as small as possible while still allowing appreciable particle counts.

\begin{figure}
\includegraphics[width=\linewidth]{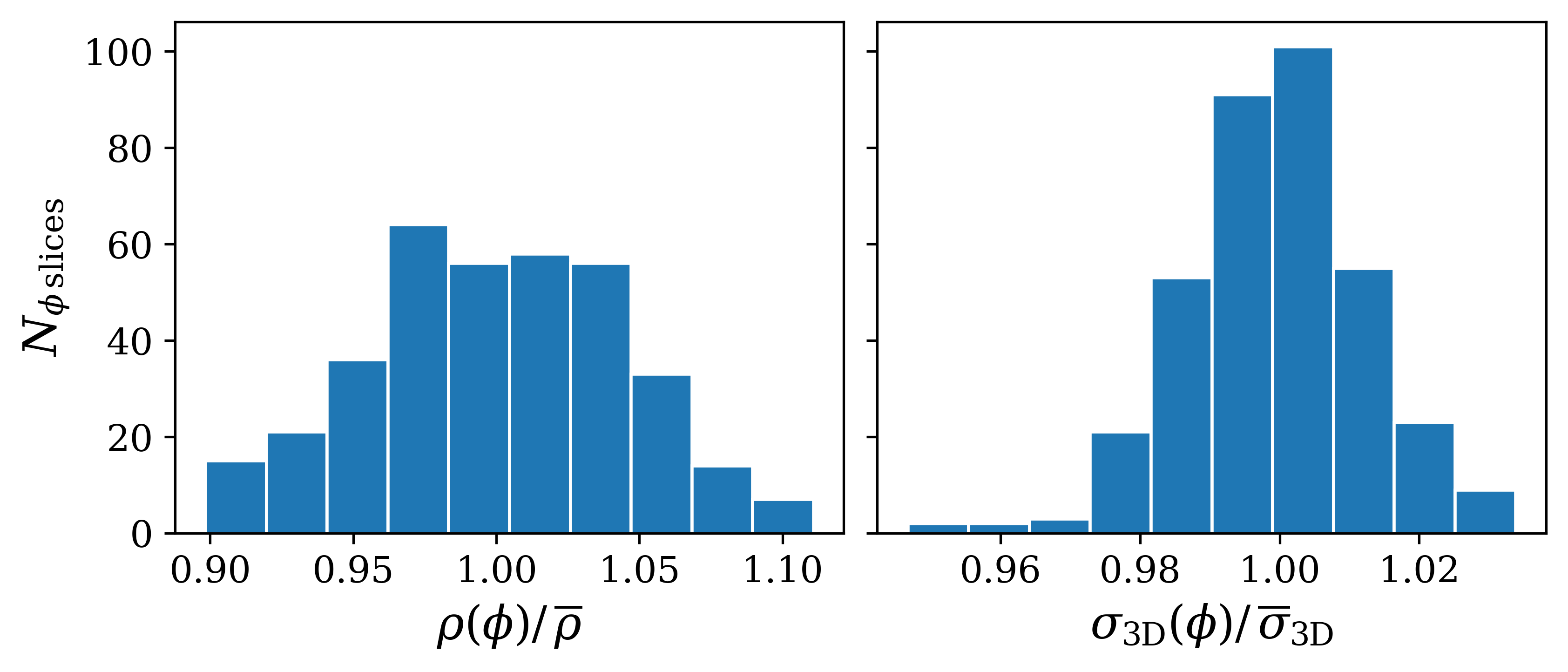}
\caption{Azimuthal variation in dark matter density (left) and velocity dispersion (right) across 30 different Sun-like regions in each of our simulated galaxies.  In both plots, measurements were made in 30 evenly-divided regions in each galaxy's solar ring and then normalized by the average density ($\overline{\rho}$) or velocity dispersion ($\overline{\sigma}_{\rm3D}$) over the full ring.    Across all galaxies, the standard deviation of $\rho(\phi)/\overline{\rho}$ is \data{stdev_linear_dendiff}, and the standard deviation of $\sigma_{\rm3D}(\phi)/\overline{\sigma}_{\rm3D}$ is  \data{stdev_linear_dispdiff}.}
\label{fig:disc_diffs}
\end{figure}

\subsection{Circular Speed}
\label{sec:vc}
All the models in this study make predictions based on a sliding scale of one simple observable: the galaxy's observed circular rotation speed at the solar location ($R_0=8.3$ kpc). In an idealized, spherically-symmetric system, $v_\mathrm{c}=\sqrt{GM/R_0}$. However, in reality, the Milky Way is a flattened disk, and material in such a disk will not rotate at this idealized speed \cite{binney2008galactic}.
We determine the circular speeds of our galaxies as the average azimuthal velocity component $\vec{v}\cdot\hat{\phi}$ of gas colder than $10^4$ K in the solar ring, where $\hat{\phi}$ points in the direction of the galaxy's rotation. This temperature ensures that the observed gas is in circular motion, because the implied ${\sim}16\rm\,\velunit$ 3D velocity dispersion is an order of magnitude smaller than $v_{\rm c}$. Note that we find that these ``observed" circular speeds end up being very close to what we would have obtained with the idealized spherical approximation. For example, Figure \ref{fig:gmr_vs_vc} in the Appendix plots our mock ``observed" circular speeds versus the idealized spherical approximation and shows that they obey close to a one-to-one relationship.
Additionally, ref. \cite{sands2024rotation} found that, for Milky Way size galaxies in FIRE, observational methods exhibit excellent agreement with the true rotation curve. Even the three largest m11's have observable rotation curves that match the true curve to within 10\%.
This suggests that our results will be largely insensitive to how we define $v_\mathrm{c}$ in the simulations. 

Ultimately, our goal is to use our inferred relationships between  simulated dark matter properties and $v_{\rm c}$ on the Milky Way itself. In doing so, we adopt $v_\mathrm{c,MW}=229\pm7\ \mathrm{km\,s^{-1}}$ for the Milky Way's rotation speed near the Sun \citep{eilers2019the-circular}.  This error bar includes systematic uncertainties provided by ref. \cite{eilers2019the-circular}. Even so, at 3\%, it contributes little to the overall uncertainty in our predictions, which is dominated by galaxy-to-galaxy variations.

\subsection{Density and Dispersion Variation within Disks}
\label{sec:den_disp}

As discussed above, we use ring-like regions centered on mock solar locations to determine the local density $\rho$, velocity dispersion $\sigma_\mathrm{3D}$, and speed distributions of dark matter particles in each simulation. It is interesting to ask if there is any significant variation in density or velocity dispersion as we move around the plane of the disks at fixed radii $R_0$. 

In order to explore this, we divided each ring into 30 $\phi$ slices and calculated $\rho(\phi)$ and $\sigma_\mathrm{3D}(\phi)$ in each piece.  
Figure \ref{fig:disc_diffs} shows histograms of those measurements at fixed $\phi$ divided by the corresponding galaxy's average for the whole ring.  The ring's average density is the sum of all dark matter particle masses therein divided by the volume of the ring. Its 3D velocity dispersion is the quadratic sum of the three velocity components' standard deviations, again calculating over all dark matter particles in the ring. Measures for the slices are the same but limited to the volume of the given slice. Density variation is shown on the left of the figure, and velocity dispersion variation on the right. The standard deviation of a $\phi$ slice from the aggregate solar ring is \data{stdev_linear_dendiff} for $\rho$ and \data{stdev_linear_dispdiff} for $\sigma_\mathrm{3D}$. This study therefore adopts the aggregate solar ring values as its simulated measurements, so $\rho=\overline{\rho}$, and $\sigma_{\rm3D}=\overline{\sigma}_{\rm3D}$ throughout the text and in Figures \ref{fig:den_vs_v} and \ref{fig:disp_vs_v}. We conservatively add the standard deviations in quadrature to the other uncertainties in predicting the Milky Way's local dark matter density and velocity dispersion in what follows. Note that this level of variation is quite small and adds an almost negligible amount to the error bar around the prediction.
Moreover, the number of dark matter particles in our twelve FIRE-2 solar rings ranges from 15,881 for m12r to 51,785 for Romulus.
With 30 slices, this implies
shot noise at the ${\sim}2$--$4\%$ level---not far from the standard deviation we measure directly among density slices.  This suggests that our reported variation of \data{stdev_linear_dendiff} in density is a conservative upper limit. 

We compared shot noise to the standard deviation of $\rho(\phi)/\overline{\rho}$ to determine the optimal number of slices. With more than ${\sim}30$ slices, shot noise exceeds the standard deviation between slices. In other words, counting statistics could be driving apparent azimuthal variations. We decreased the number of slices in multiples of 5 until standard deviation exceeded shot noise for all galaxies, which occurs at 30. 
Also note that our results do not appreciably change if we use an extreme number of slices. Using 200 for example, the resulting 8.9\% and 3.0\% standard deviations in density and dispersion remain nearly negligible contributors to overall uncertainty in Milky Way predictions. The main contributor to said uncertainty is galaxy-to-galaxy variation.

\subsection{Defining Escape Speed}
\label{sec:vesc}
As discussed above, it is common to truncate an assumed dark matter particle speed distribution above some ``escape speed" $v_{\rm esc}$, but knowing the precise value of the escape speed to use is non-trivial.  Even if the full potential is known, the formal escape speed depends on the distance\footnote{Qualitatively, in order for a dark matter particle to ``escape" the Solar region, it does not need to escape to infinity, but just how far is not entirely clear.} a particle needs to travel in order to ``escape" the region of interest (e.g. \citep{piffl2014the-rave, deason2019the-local}).

In what follows, we take a practical approach and define the simulated escape speed to be the value of $v$ where the local dark matter speed distribution falls to zero. To determine the sensitivity of this measure to resolution and Poisson noise, we tested the variability of the results by sampling only 10\% of each galaxy's particles and repeating the analysis ten times. The standard deviation between each galaxy's ten trials ranges from $\data{vesc_min_std}{\sim}\data{vesc_max_std}$. Therefore, the study's measure of escape speed is acceptably robust.
We also compare our ``measured" escape speed to the formal escape speed to infinity: $v_\mathrm{esc}(\Phi)=\sqrt{2|\Phi|}$. For a given galaxy, we calculate this at eight points in the solar ring, equidistant in azimuthal angle $\phi$ and take the average of those eight values. 
As shown below, this categorically overshoots the measured $v_{\rm esc}$ in our simulations, which is consistent with previous findings (e.g. \cite{deason2019the-local}). 

\subsection{Speed Distribution}
\label{sec:mcmc}
We employ a Markov chain Monte Carlo method of fitting speed distribution parameters, using the package \textsc{emcee} \citep{foreman-mackey2013emcee:}. The log-likelihood is
\begin{align}
\ln\mathcal{L}(\theta)&=-\chi^2(\theta)/2\ \mathrm{with}\\
\chi^2(\theta)&=\sum_{i=1}^{600}\frac{\left[p_v(x_i) - \hat{p}_v(x_i, \theta)\right]^2}{\sigma_i^2}.
\end{align}
where $p_v$ is the probability density of a dark matter particle in a given galaxy exhibiting speed $v$, as determined by a histogram of 50 bins ranging from 0 $\rm km\,s^{-1}$ to the speed of the fastest dark matter particle in the galaxy's solar ring. The 50 measurements from each of the 12 FIRE-2 discs yield an aggregate of 600 data points. The predicted probability density $\hat{p}_v(x_i, \theta)=4\pi v^2 f(x_i, \theta)$ depends on $\theta$, a vector of parameters $(d, e, h, j, k)$ that we optimize for the given speed distribution $f$. Section \ref{sec:distribs} describes the parameters' meanings. We assume flat priors of $d\in[100, 130]\,\rm km\,s^{-1}$, $e\in[0.8, 1.2]$, $h\in[200, 400]\,\rm km\,s^{-1}$, $j\in[0.1, 0.8]$, and $k\in[0.010, 0.045]$. The feature matrix $x$ consists of columns 1) dark matter particle speed corresponding to the midpoints of the histogram bins and 2) circular speed at the solar location $v_\mathrm{c}=\langle\vec{v}\cdot\hat{\phi}\rangle_{T\leq 10^4\mathrm{K}}$ discussed in Section \ref{sec:vc}. For a given galaxy's entries into the circular-speed column of the feature matrix, the same $v_\mathrm{c}$ value is repeated 50 times. We use Poisson errors so $\sigma_i = p_v(x_i)/\sqrt{N_i}$ where $N_i$ is the number of particles in the given speed bin.

\section{Results}
\label{sec:results}

\begin{figure}
\centering
\includegraphics[width=\textwidth]{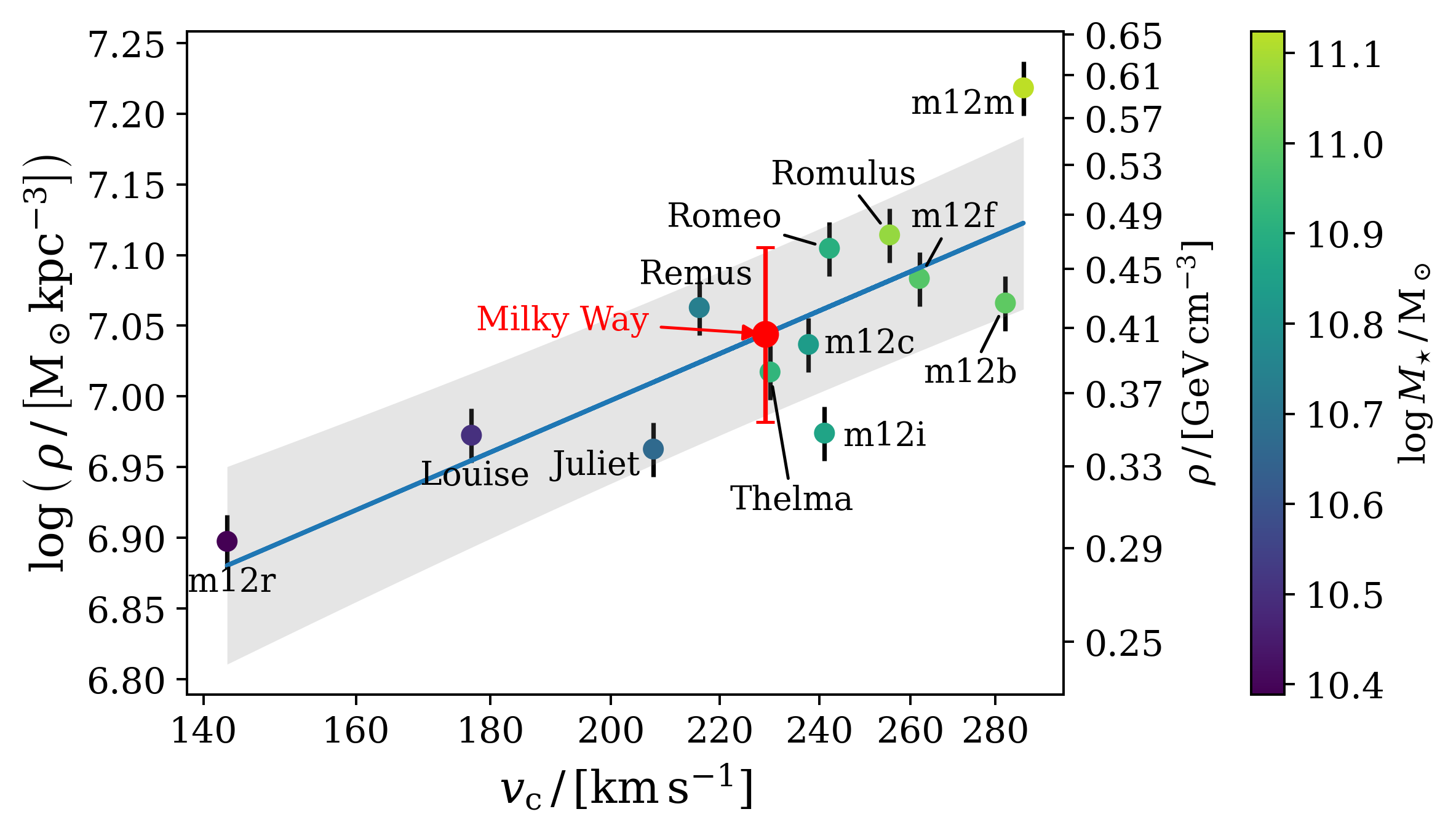}
\caption{Relationship between the local dark matter density and circular speed for our simulated galaxies.  The name of each simulated galaxy (as defined in the original FIRE-2 papers) is provided for each respective point.  The color bar maps to the total stellar mass of each galaxy for reference. The blue line shows the best fit from Equation \ref{eqn:rho}, which yields $\rho \sim v_{\rm c}^{\data{den_slope}}$ and a log-space coefficient of determination $r^2$ of $0.65$. The black error bars on each data point come from our measured azimuthal deviations in local density, discussed in Section \ref{sec:den_disp}. The grey band represents the 1$\sigma$ uncertainty around the prediction line. The red point and error bar represents our Milky Way prediction: $\rho(v_\mathrm{c,MW})=\data{rho_1e7msun}^{+\data{drho_1e7msun_plus}}_{-\data{drho_1e7msun_minus}}\,10^7\mathrm{M_\odot\,kpc^{-3}}=\data{rho_GeV}\pm\data{drho_GeV}\,\denunit$.
}
\label{fig:den_vs_v}
\end{figure}

\subsection{Local Dark Matter Density}
Figure \ref{fig:den_vs_v} shows the relationship between the measured ``local'' dark matter densities at mock solar locations in our simulations versus the local circular speed. We find that the relationship between density and $v_\mathrm{c}$ is well described by a power law:
\begin{equation}
\rho=\rho_0\left(\frac{v_\mathrm{c}}{100\,\mathrm{km\,s^{-1}}}\right)^{\alpha}, 
\label{eqn:rho}
\end{equation}
where $\rho_0=\data{rho0_1e7msun}^{+\data{drho0_1e7msun_plus}}_{-\data{drho0_1e7msun_minus}}10^7\mathrm{M_\odot\,kpc^{-3}}=\data{rho0_GeV}^{+\data{drho0_GeV_plus}}_{-\data{drho0_GeV_minus}}\denunit$ and $\alpha=\data{den_slope}\pm\data{dden_slope}$.
Section \ref{sec:errors} in the Appendix discusses our treatment of uncertainties in the model.
The solid blue line in Figure \ref{fig:den_vs_v} visualizes this density model, which exhibits  an $r^2$ coefficient of determination of 0.65. Note that the coefficient of determination is calculated based on log densities. The same applies to the models that follow below for velocity dispersion and escape speed. The shaded band shows the one-sigma region, and the color bar maps to stellar mass of the galaxy. 

If we apply our model fit to the Milky Way, assuming a circular speed of $v_\mathrm{c,MW}=229\pm7\ \mathrm{km\,s^{-1}}$, from ref. \cite{eilers2019the-circular}, this yields
\begin{align}
\rho(v_\mathrm{c,MW})&=\data{rho_1e7msun}^{+\data{drho_1e7msun_plus}}_{-\data{drho_1e7msun_minus}}10^7\mathrm{M_\odot\,kpc^{-3}}\label{eqn:rho_mw_astro}\\
&=\data{rho_GeV}\pm\data{drho_GeV}\,\denunit.\label{eqn:rho_mw_particle}
\end{align}
The quoted error includes the uncertainty from the fit as well as allowance for variance within the disk. The latter is equal to the azimuthal variation found in our simulations in Section \ref{sec:den_disp}. 
Interestingly, our determination of the local density is quite close to the aggregate values in Table \ref{tab:rho}. Not only does this cross-validate this study's density result with others', it also supports using these select FIRE-2 galaxies for making other Milky Way predictions. Note that one could easily adopt a different value for $v_\mathrm{c,MW}$ through Equation \ref{eqn:rho}. When making predictions around $v_\mathrm{c}\approx229\,\mathrm{km\,s^{-1}}$, one can assume the corresponding uncertainties are the same as those we cite.

\begin{figure}
\includegraphics[width=\textwidth]{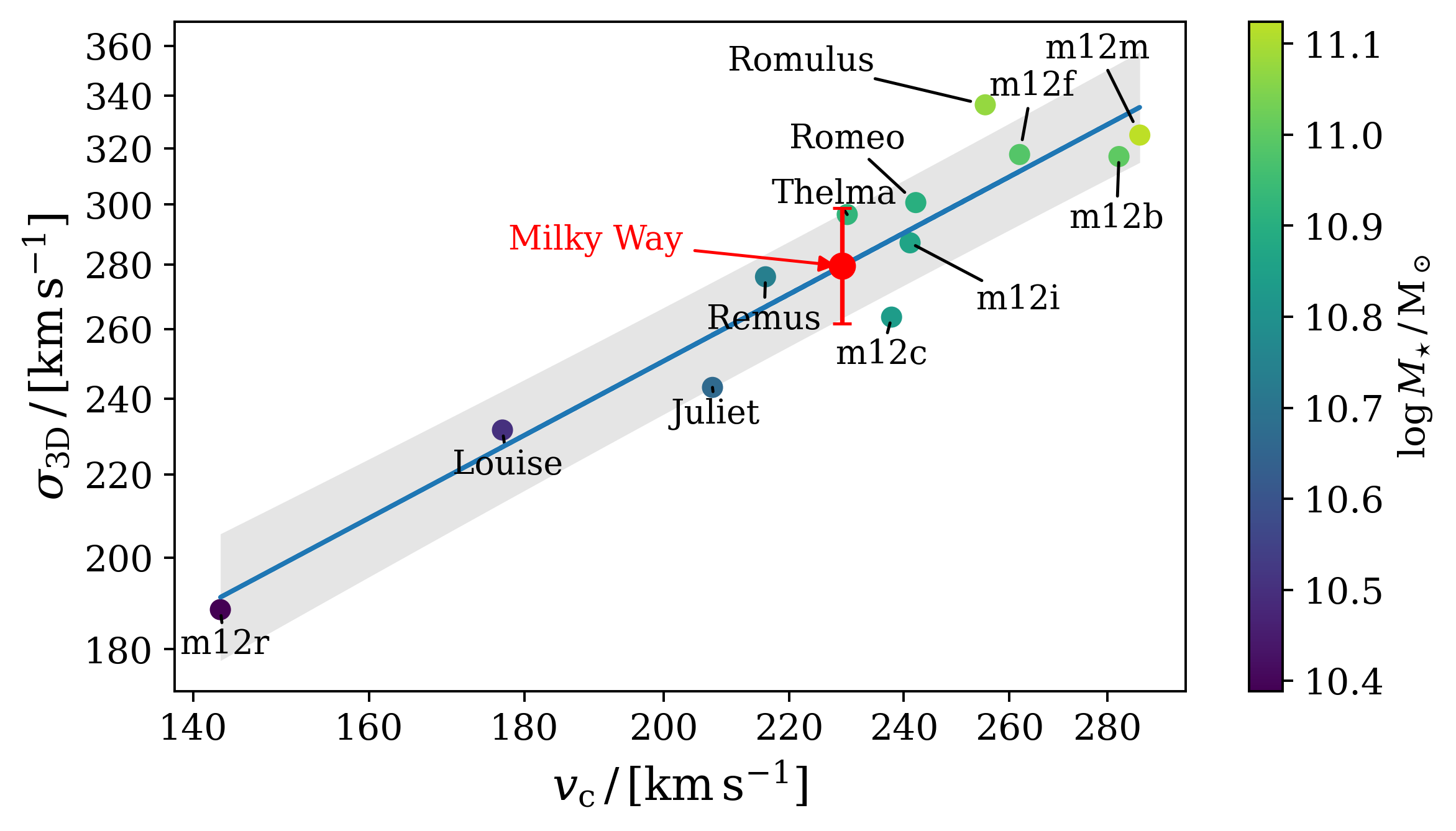}
\centering
\caption{This is the same as Figure \ref{fig:den_vs_v} but now showing measured velocity dispersion versus circular speed for each simulated galaxy. The best-fit model (Equation \ref{eqn:sigma_vs_vc}, blue line) has $\sigma \sim v_{\rm c}^{\data{disp_slope}}$ and  a log-space coefficient of determination $r^2$ of 0.90. The relationship between local dark matter velocity dispersion and local circular speed is even tighter than that between the latter and density shown in Figure \ref{fig:den_vs_v}. The red point and error bar shows our Milky Way prediction: $\sigma_{\rm 3D}(v_\mathrm{c,MW})=\data{disp}^{+\data{ddisp_plus}}_{-\data{ddisp_minus}}\,\mathrm{km\,s^{-1}}$.} 
\label{fig:disp_vs_v}
\end{figure}

\subsection{Local Dark Matter Velocity Dispersion}
We find that the dark matter velocity dispersion is even more tightly correlated with $v_\mathrm{c}$ than the density is, with 
\begin{equation}
\sigma_\mathrm{3D}=\sigma_0\left(\frac{v_\mathrm{c}}{100\,\mathrm{km\,s^{-1}}}\right)^{\gamma},
\label{eqn:sigma_vs_vc}
\end{equation} 
where $\sigma_0=\data{disp_amp}^{+\data{ddisp_amp_plus}}_{-\data{ddisp_amp_minus}}\,\velunit$, and $\gamma=\data{disp_slope}\pm\data{ddisp_slope}$, which exhibits a 0.90  $r^2$ coefficient of determination. For the Milky Way, this yields 
\begin{equation}
\sigma_\mathrm{3D}(v_\mathrm{c,MW})=\data{disp}^{+\data{ddisp_plus}}_{-\data{ddisp_minus}}\,\mathrm{km\,s^{-1}}.
\label{eqn:sigma_val}
\end{equation}
Figure \ref{fig:disp_vs_v} visualizes the model and its Milky Way prediction versus the data. Similarly to our density model, one could adopt a different MW circular speed by simply plugging their preferred value into Equation \ref{eqn:sigma_vs_vc}.

\begin{figure}
\centering
\includegraphics[width=\textwidth]{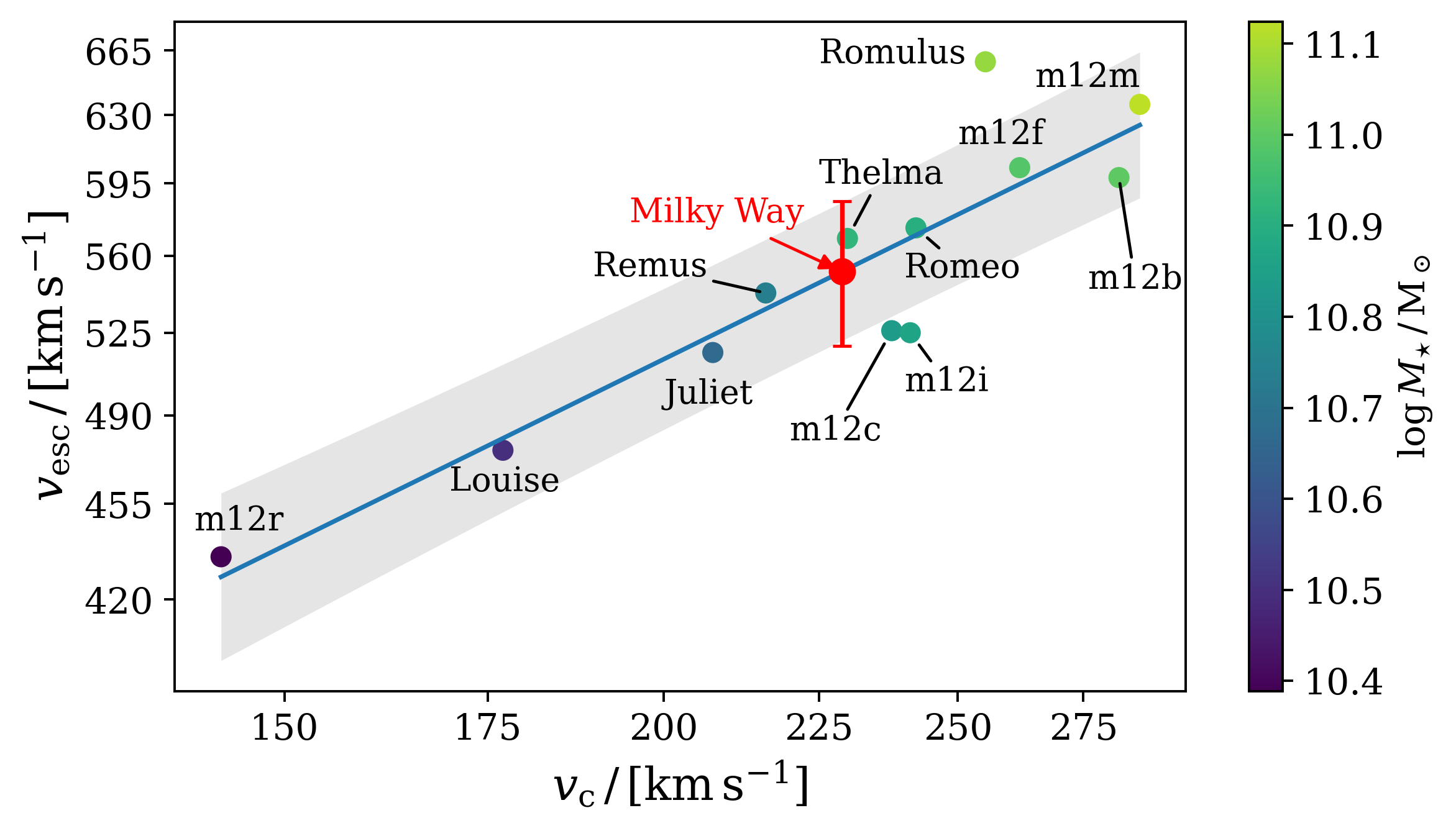}
\caption{The escape speed, $v_{\rm esc}$, versus circular speed for each simulated galaxy. Here, the escape speed is set to be the speed of the fastest dark matter particle in the given simulation's solar ring. The blue line shows the power-law model from Equation \ref{eqn:vesc(vc)}, $v_{\rm esc} \sim v_c^{\data{veschat_slope}}$, which has a 0.82 $r^2$ coefficient of determination in log space. The grey band shows the 1$\sigma$ uncertainty around the prediction. The red point with error bars shows our prediction for the local escape velocity near the Sun in the Milky Way using this model: $v_{\rm esc}(v_\mathrm{c,MW})=\data{vesc_mw(vc)}\pm\data{dvesc_mw(vc)}\,\velunit$.} 
\label{fig:vlim}
\end{figure} 

\begin{figure*}[!t]
\includegraphics[width=\textwidth]{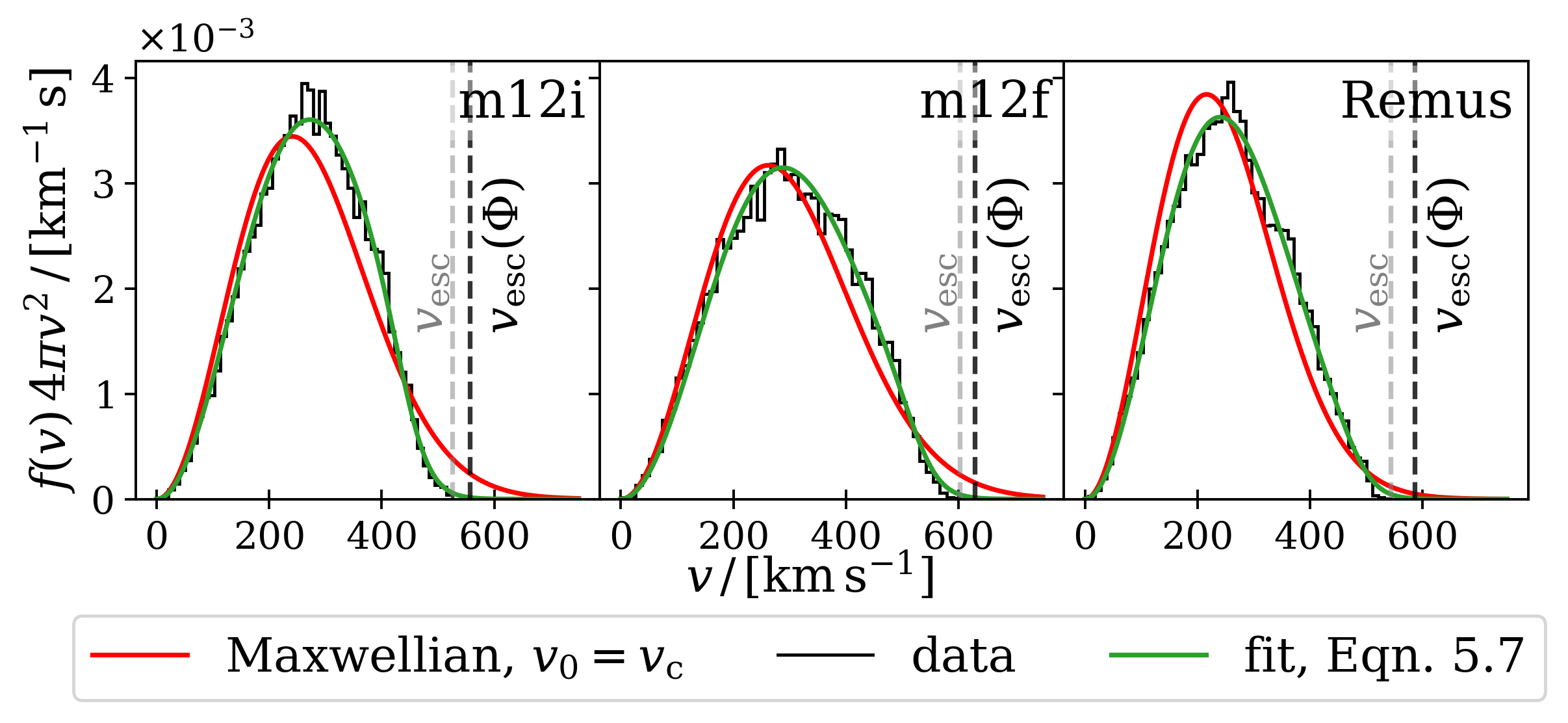}
\caption{This figure shows the measured speed distribution data for three example systems in black histograms. The \maxwellcolor{} lines show the Maxwellian distributions defined by Equation \ref{eqn:simple_max}, which have only one parameter, the peak speed $v_0$. In this figure, we adopt the common assumption that peak speed is equal to the circular speed $v_\mathrm{c}$ measured in each galaxy. This assumption under-predicts the peak speed.
For illustrative purposes, the grey vertical dashed lines show the escape speed, $v_{\rm esc}$, the speed of the fastest dark matter particle in a given simulation's solar ring.
The plots here also show the escape speed based on potential, $v_{\rm esc}(\Phi)$, where we take escape distance to be infinity. Section \ref{sec:vesc} provides further detail on the calculation method for $v_{\rm esc}(\Phi)$. Take notice of the difference between these two possible cutoff speeds. If one were to cut off the distributions at the escape speeds determined by potential, one would assume the presence of too many high-speed particles.
However, even if we were to cut the distributions at the $v_{\rm esc}$ we measure, the Maxwellian would still give too many high-speed particles. 
The green lines show best fits for each individual galaxy using the two-parameter function given by Equation \ref{eqn:final_model}. The green fits to the black data are strikingly tight and demonstrate that this functional form can, in principle, reproduce the speed distributions accurately.}
\label{fig:maxwellian}
\end{figure*}

\subsection{Escape Speed}
As discussed in Section \ref{sec:vesc}, we define the escape speed, $v_\mathrm{esc}$, as the smallest speed above which we find no dark matter particles in the simulated local region. Figure \ref{fig:vlim} shows the relationship between $v_\mathrm{esc}$ measured this way and the local circular speed. The two parameters correlate as
\begin{equation}
v_{\rm esc}(v_{\rm c})=v_{\rm e,0}\left(\frac{v_\mathrm{c}}{100\,\velunit}\right)^\varepsilon,
\label{eqn:vesc(vc)}
\end{equation}
where $v_{\rm e,0}=\data{veschat_amp}^{+\data{dveschat_amp_plus}}_{-\data{dveschat_amp_minus}}\velunit$ and $\varepsilon=\data{veschat_slope}\pm\data{dveschat_slope}$. Its predictions exhibit a 0.82  $r^2$ coefficient of determination. Figure \ref{fig:vlim} visualizes this fit and shows our $v_\mathrm{esc}(v_\mathrm{c,MW})=\data{vesc_mw(vc)}\pm\data{dvesc_mw(vc)}\,\velunit$ estimate for the Milky Way. As we continue to emphasize, this and the other models in this paper are calibrated to  simulation data that would not be observable in the real world but, once fit, require only one single observable feature---the local circular speed---to predict those non-observable features of the Milky Way. 

\subsection{Speed Distribution}
\label{sec:distribs}
A common assumption for the dark matter speed distribution is a simple Maxwellian (Equation \ref{eqn:simple_max}).  
Figure \ref{fig:maxwellian} compares the true speed distribution data for three of our galaxies (black) to the Maxwellian assumption (\maxwellcolor{}), where we have fixed the peak speed, $v_0$, in  Equation \ref{eqn:simple_max} to the local circular speed, $v_c$, as is the standard choice.  We see that the Maxwellian curve peaks at too low of a speed and has an excess high-speed tail. Even if we truncated the Maxwellian distribution at the formal escape speed set by the potential (dashed black line) or the escape speed calculated by our approach (Section \ref{sec:vesc}, grey dashed line), the Maxwellian curve would still over-produce counts at high speeds. We find that similar shape mismatches exist for all of our simulated galaxies.

\begin{figure*}
\includegraphics[width=\textwidth]{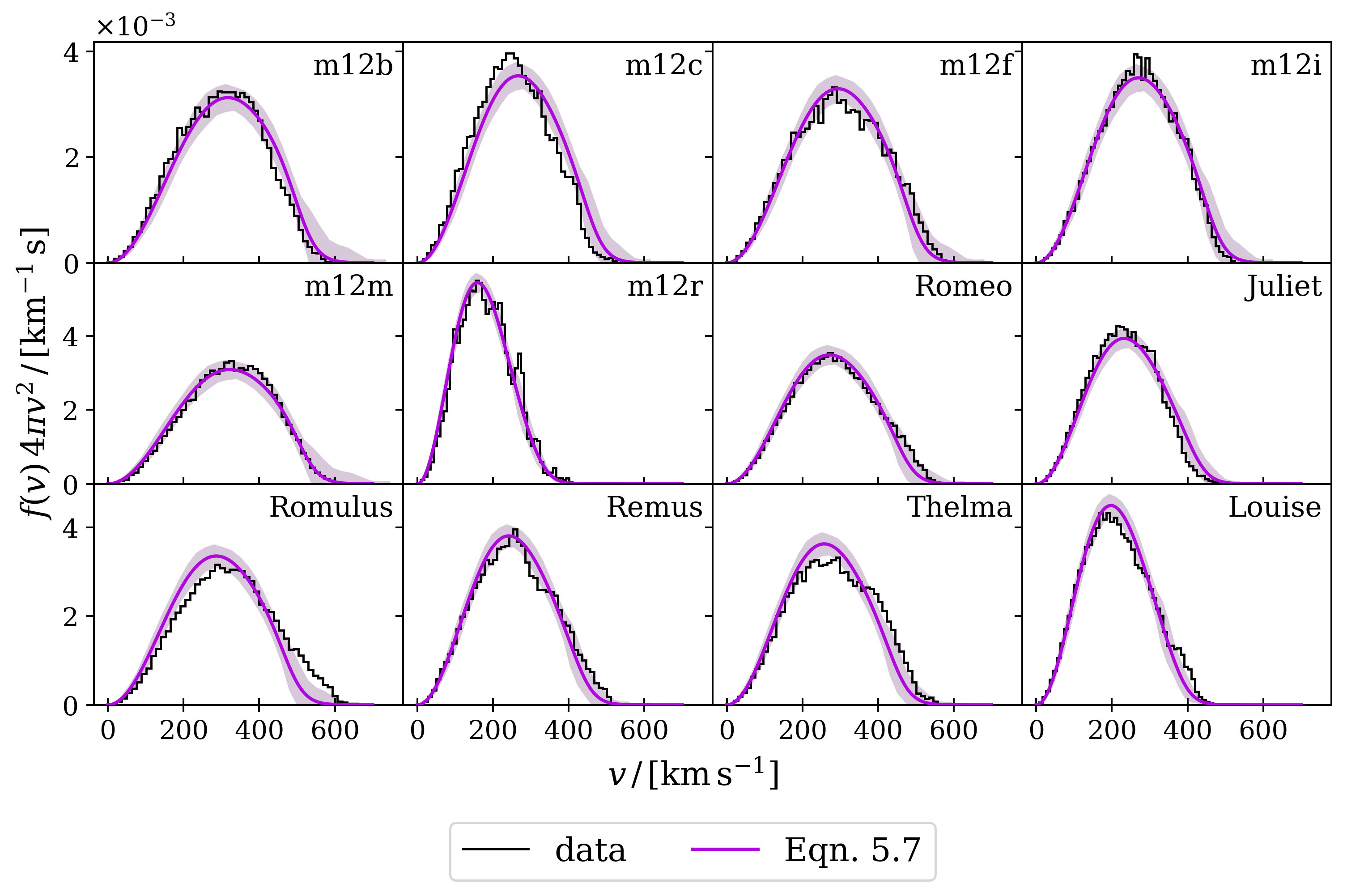}
\caption{The \staudtcolor{} lines represent the best-fit speed distributions predicted by this study's final model using only the galaxy's $v_{\rm c}$, specifically using Equations \ref{eqn:final_model}, \ref{eqn:final_v0}, and \ref{eqn:final_vdamp}.  The \distribbandcolor{} bands represent the uncertainty around those predictions. Table \ref{tab:errors} provides the values of the uncertainty band, which are the root mean squares of the deviations between the prediction and the data at a given $v/v_0(v_{\rm c})$.}
\label{fig:universal_fits}
\end{figure*}

While the Maxwellian curves shown in Figure \ref{fig:maxwellian} are not best-fit Maxwellian shapes to each galaxy,  we have found that even allowing for a variable $v_0$ for each galaxy, the simple Maxwellian fit fails. As discussed in the introduction, others have found similar results and tried to provide better descriptions of the speed distribution by employing models such as generalized Tsallis distributions \citep{ling2010dark, lacroix2020predicting, bozorgnia2016simulated, nunez-castineyra2019dark} or other modifications to the Maxwellian distribution \citep{mao2013halo-to-halo, lisanti2011dark}. 

We find a good fit by introducing a speed parameter $v_\mathrm{damp}$, above which each distribution is suppressed:
\begin{equation}
f(v)=\frac{1}{N(v_0,v_\mathrm{damp})}\exp\left(-\frac{v^2}{v_0^{2}}\right)
\mathcal{S}(v_\mathrm{damp}-v),
\label{eqn:final_model}
\end{equation}
where $\mathcal{S}$ is the sigmoid function
\begin{equation}
\mathcal{S}(\Delta v)=\frac{1}{1+\exp(-k\Delta v)},
\label{eqn:theta}
\end{equation}
and $k$ represents how strongly the sigmoid suppresses the distribution around $v_\mathrm{damp}$. We adopt $k=\data{k}$  because it is the best-fit value for the final universal model at the end of this section.
The green lines in Figure \ref{fig:maxwellian} plot this model using the best fit parameter set $(v_0, v_\mathrm{damp})$ for each individual galaxy; it matches the data remarkably well. We find similar success for all of the galaxies in our simulated sample. Note that these fits use the non-observable dark matter speed distributions to directly determine $(v_0, v_\mathrm{damp})$. Such a fit does not yet gauge our ability to define the best pair of parameters to use for the Milky Way itself; however, it shows that, in principle, the model's shape has enough flexibility to reproduce the range of distributions we predict.

Our goal now is to determine the best pair of parameters $(v_0, v_\mathrm{damp})$ to apply to the Milky Way. Our approach is to search for a way to predict each simulated galaxy's values using power law models on their local circular speed: 
\begin{equation}
v_0=d\left(\frac{v_\mathrm{c}}{100\,\mathrm{km\,s^{-1}}}\right)^e
\label{eqn:final_v0}
\end{equation}
and
\begin{equation}
v_\mathrm{damp}=h\left(\frac{v_\mathrm{c}}{100\,\mathrm{km\,s^{-1}}}\right)^j.
\label{eqn:final_vdamp}
\end{equation}
We find $d=\data{d}\pm\data{dd}\,\mathrm{km\,s^{-1}}$, $e=\data{e}\pm\data{de}$, $h=\data{h}\pm\data{dh}\ \mathrm{km\,s^{-1}}$, and $j=\data{j}\pm\data{dj}$ provides a good description for our set of twelve model galaxies. We fit all these parameters, including $k$, by running the Markov chain Monte Carlo package \textsc{emcee} \citep{foreman-mackey2013emcee:} on a concatenation of all simulated galaxies.  In this process we find the best single value of $k$ for the concatenation, and this results in the adoption of $k=\data{k}\pm\data{dk}$ mentioned earlier.  The individual galaxy fits shown in green in Figure \ref{fig:maxwellian} use the same $k$ as the one that optimizes this universal model. 

Figure \ref{fig:universal_fits} shows the speed distribution data for all twelve Milky Way sized FIRE-2 disks (black histograms) and our model's predictions based only on each galaxy's $v_c$ (\staudtcolor{} lines). The \distribbandcolor{} bands represent the total uncertainty.
The statistical portion thereof, related to the error bars on the model parameters, is extraordinarily small. 
When plotted, its width is hidden behind the \staudtcolor{} prediction line in Figure \ref{fig:universal_fits}. This indicates that the majority of the uncertainty between our model and the data is systematic. Therefore, we determine the total uncertainty band as follows. 
We divide each $x$-axis by that galaxy's $v_0(v_{\rm c})$, given by Equation \ref{eqn:final_v0}, and calculate the residuals between the \staudtcolor{} predictions and the black data. There are then 12 residuals, one for each galaxy, at every $v/v_0$. The root mean square of each set of 12 gives the total uncertainty in $4\pi v^2 f(v)$ for a given $v/v_0$. We provide these numerical uncertainties in Table \ref{tab:errors}. This simple model does a good job of predicting the dark matter speed distributions in each simulated galaxy using only $v_c$. 

\begin{table}[t]
\centering
\caption{Total uncertainty in $4\pi v^2 f(v)$ where $f(v)$ is this study's speed distribution model given by Equations \ref{eqn:final_model}--\ref{eqn:final_vdamp}. We calculate it as the root
mean square of the deviations between the prediction and the simulation data at a given $v/v_0(v_{\rm c})$ and provide it below in units of $10^{-3}\,\rm km^{-1}s$. These values yield the \distribbandcolor{} bands in Figures \ref{fig:universal_fits} and \ref{fig:distrib_mw}.}
\label{tab:errors}

\begin{filecontents*}{tables/errors.csv}
0.04,0.01
0.12,0.03
0.19,0.07
0.27,0.10
0.35,0.18
0.43,0.22
0.51,0.23
0.58,0.22
0.66,0.24
0.74,0.27
0.82,0.29
0.90,0.25
0.97,0.24
1.05,0.21
1.13,0.23
1.21,0.25
1.29,0.28
1.36,0.28
1.44,0.26
1.52,0.25
1.60,0.29
1.68,0.49
1.75,0.47
1.83,0.34
1.91,0.30
1.99,0.26
2.07,0.17
2.14,0.07
2.22,0.04
2.30,0.05
\end{filecontents*}
\csvreader[
    no head,
    tabular = cc,
    range   = 1-10,
    table head = 
        \toprule $v/v_0$ & Uncertainty \\
        \midrule
]{tables/errors.csv}{}{\csvcoli & \csvcolii}
\hspace{5mm}
\csvreader[
    tabular = cc,
    range   = 10-19,
    table head = 
        \toprule $v/v_0$ & Uncertainty \\
        \midrule
]{tables/errors.csv}{}{\csvcoli & \csvcolii}
\hspace{5mm}
\csvreader[
    tabular = cc,
    range   = 20-,
    table head = 
        \toprule $v/v_0$ & Uncertainty \\
        \midrule
]{tables/errors.csv}{}{\csvcoli & \csvcolii}

\end{table}

\begin{figure}[t!]
\centering
\includegraphics[width=\textwidth]{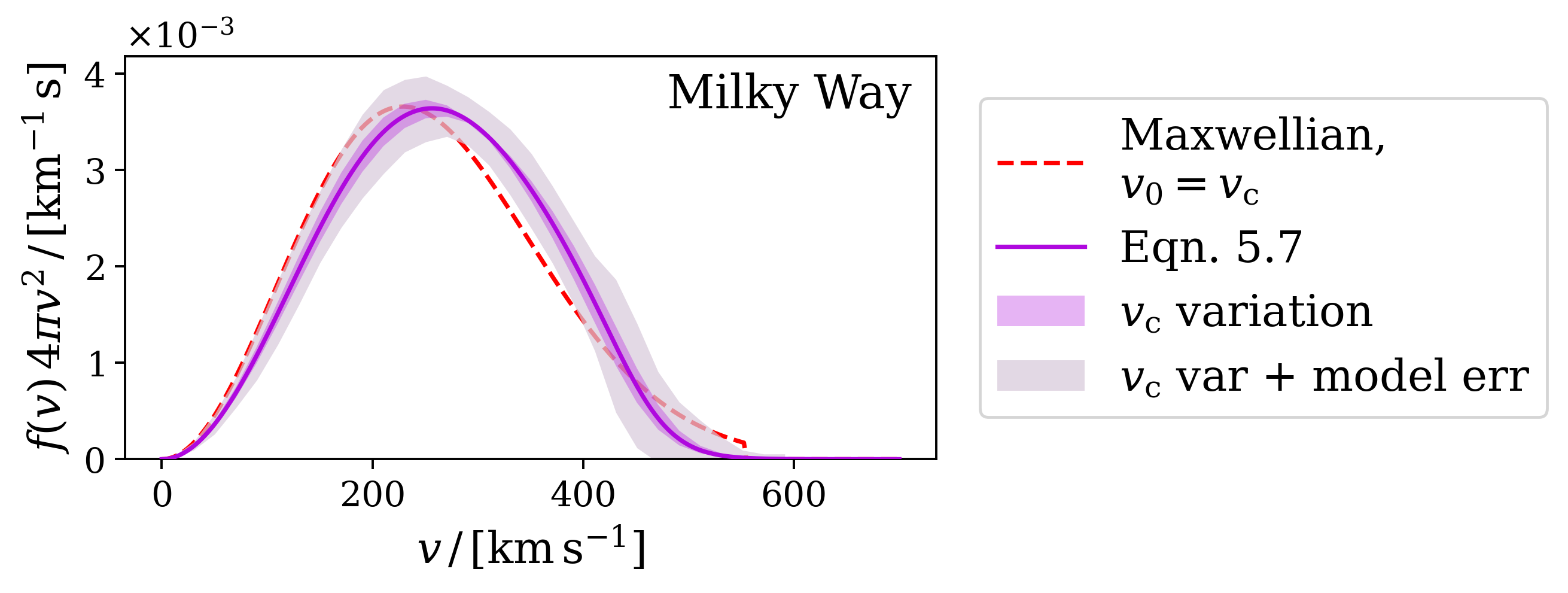}
\caption{The \staudtcolor{} line represents this work's final prediction for the Milky Way's speed distribution. The \vcvarbandcolor{} band exhibits the range of results given by circular speeds in the $\pm7\,\rm km\,s^{-1}$ error band from ref. \cite{eilers2019the-circular}. The \distribbandcolor{} band combines the $v_{\rm c, MW}$ uncertainty with the model uncertainty quantified in Table \ref{tab:errors}. The \maxwellcolor{} dashed line shows a Maxwellian with peak speed equal to the local circular speed and a sharp cutoff at the escape speed. We use Equation \ref{eqn:vesc(vc)} to determine $v_{\rm esc}(v_{\rm c, MW})=\data{vesc_mw(vc)}\,\velunit$.}
\label{fig:distrib_mw}
\end{figure}

Figure \ref{fig:distrib_mw} provides our prediction for the Milky Way's local dark matter speed distribution using $v_c = v_\mathrm{c,MW}=229\pm7\ \mathrm{km\,s^{-1}}$. The implied parameters in Equation 
\ref{eqn:final_model} are $v_0(v_{\rm c, MW})=\data{v0_mw}\pm\data{dv0_mw}\,\velunit$ and $v_{\rm damp}(v_{\rm c, MW})=\data{vdamp_mw}\pm\data{dvdamp_mw}\,\velunit$, and these generate the \staudtcolor{} prediction line. 
Given the tight margin of statistical error on the parameters $d$, $e$, $h$, and $j$, the majority of uncertainty in $v_0(v_{\rm c, MW})$ and $v_{\rm damp}(v_{\rm c, MW})$ comes from the $\pm7\,\rm km\,s^{-1}$ uncertainty in the Galaxy's circular speed \citep{eilers2019the-circular}. That said, the \distribbandcolor{} uncertainty band in the end result, the speed distribution, is dominated by the systematics of galaxy-to-galaxy variation.
The dashed \maxwellcolor{} line shows the standard Maxwellian assumption with a sharp cutoff at the escape speed. We use Equation \ref{eqn:vesc(vc)} to determine $v_{\rm esc}(v_{\rm c, MW})=\data{vesc_mw(vc)}\,\velunit$. Notice that the peak speed of our model is higher than in the standard assumption, while the damping behavior of our model pulls the high-speed tail below the Maxwellian starting around $\data{vcrit_mw}\,\velunit$. The difference between the curves points to the benefits of using this work's prediction for the Milky Way's speed distribution over the standard assumption.

\begin{figure}[!t]
\centering
\includegraphics[width=\textwidth]{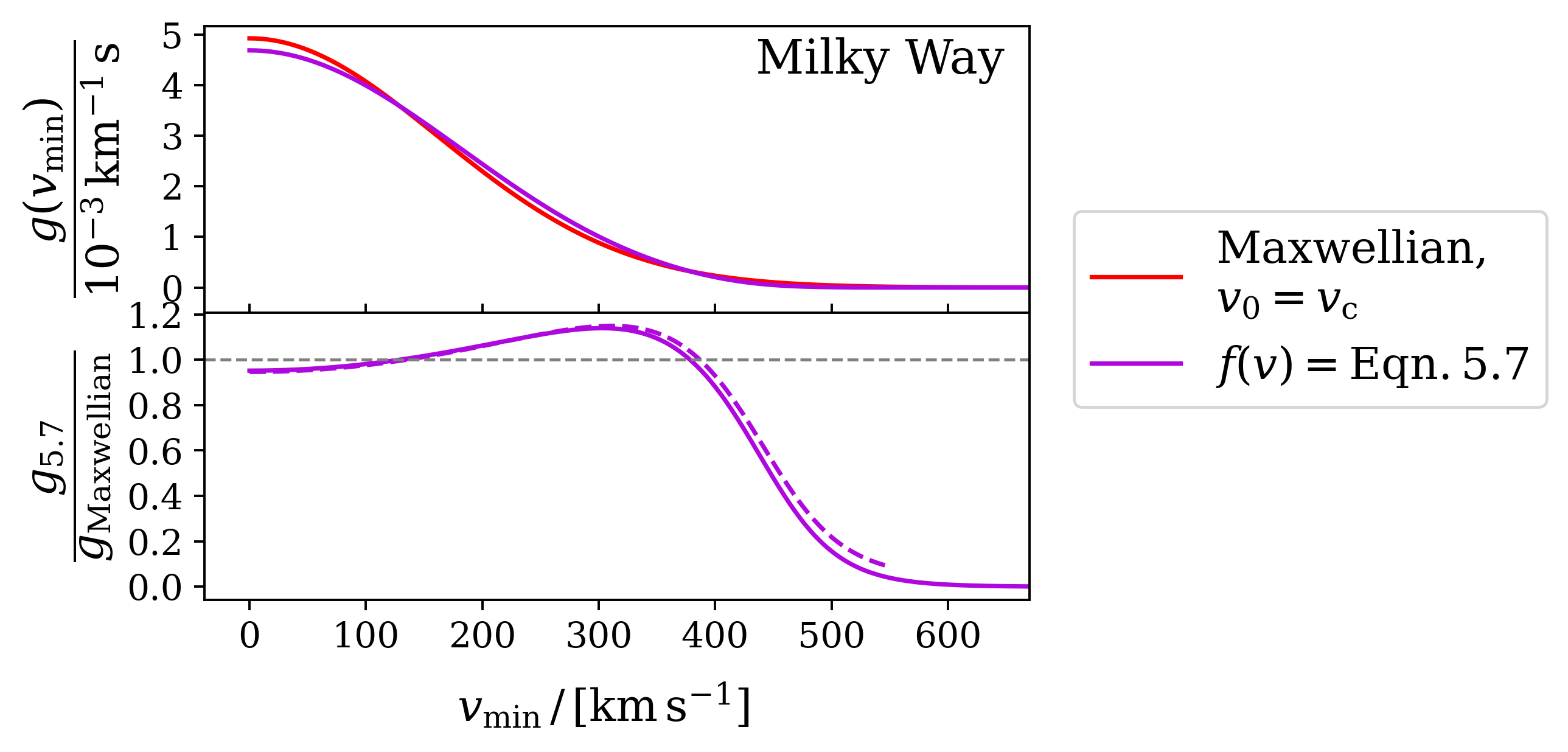}
\caption{Top: Comparison of the halo integral for the Maxwellian model in \maxwellcolor{} versus this study's model in \staudtcolor{}. Bottom: The solid \staudtcolor{} line shows the ratio of the halo integral using Equation \ref{eqn:final_model} divided by that for the Maxwellian. Given that standard practice is to cut the distribution at the escape speed, we also cut both models at $v_{\rm esc}(v_{\rm c, MW})=\data{vesc_mw(vc)}\,\velunit$ and show the ratio in a dashed line.}
\label{fig:halo_integral_mw}
\end{figure}

Figure \ref{fig:halo_integral_mw} shows the implied halo integral (Equation \ref{eqn:halo_integral}) for our predicted speed distribution in the Milky Way along with the same for the standard Maxwellian assumption. Comparing the two, the Maxwellian over-predicts the halo integral at low minimum speeds, owing to its lower peak speed. It under-predicts in the intermediate range and returns to an excess at the high end.  Given that standard practice is to truncate the Maxwellian at the escape speed, the dashed line in the lower panel does that for both models, using $v_{\rm esc}(v_{\rm c, MW})=\data{vesc_mw(vc)}\,\velunit$, our best-fit escape speed for the Milky Way. This improves the Maxwellian slightly at high speeds. However, the divergence remains.

\section{Discussion and Conclusions}
We have used FIRE-2 simulations of Milky Way size disk galaxies to inform predictions of the local dark matter density, velocity dispersion, and speed distribution near the Sun using one observable: the circular speed near the Sun, $v_c$.  We then use our best-fit models to inform dark matter expectations in the Milky Way using $v_{\rm c,MW}=229\pm7\,\velunit$, as determined by ref. \cite{eilers2019the-circular}.   Our main results are summarized in Table \ref{tab:params}. All these parameters are crucial for informing and interpreting  dark matter direct detection experiments. 

\textbf{Local dark matter density:} In our simulations, the local dark matter density near solar locations is well characterized by a power law $\rho \propto v_{\rm c}^{\alpha}$, with $\alpha \simeq 0.8$ (Equation \ref{eqn:rho} and Figure \ref{fig:den_vs_v}).  How does this scaling compare to naive expectations? For a power-law density profile in total mass, $\rho_{\rm tot} \propto 1/r^n$, we would expect $\rho_{\rm dm}(R_0) \propto \mathcal{F} v_{\rm c}^2 (3-n)$, where $\mathcal{F}$ is the fraction of the total mass in dark matter; $\rho_{\rm dm} = \mathcal{F} \rho_{\rm tot}$.  In our simulations, $\mathcal{F} \propto 1/v_{\rm c}^{1.3}$ if we measure $\mathcal{F}$ in the solar ring; this would imply $\rho_{\rm dm}(R_0) \propto  v_{\rm c}^{0.7} (3-n)$, where the power on $v_{\rm c}$ is close to the 0.8 this study finds in the FIRE-2 simulations. 

Assuming that the same model holds for the Solar location in the Milky Way, we find $\rho(v_{\rm c, MW})=\data{rho_GeV}\pm\data{drho_GeV}\,\denunit$.  This value is similar to those obtained by more complicated kinematic models and is remarkably close to the average of past estimates since 2018 summarized in Table \ref{tab:rho}. Another relevant comparison is to ref. \cite{sofue2020rotation}, who find $\data{rho_canon_GeV}\pm\data{drho_canon_GeV}\,\denunit$ using the average of 
15 independent $\rho$ determinations from literature published between 2010 and 2020.

\textbf{Local dark matter velocity dispersion:}  We find that the 3D velocity dispersion of dark matter near mock solar locations in our simulations follows a power law given by Equation \ref{eqn:sigma_vs_vc} (see Figure \ref{fig:disp_vs_v}).  Note that the best-fit power-law is slightly flatter than linear; $\sigma \propto v_{\rm c}^{\gamma}$, with $\gamma \simeq 0.8$.  This is different than what is expected in an isothermal sphere model, where the 3D velocity dispersion is linear with circular speed as $\sigma_{\rm 3D} = \sqrt{3/2} \, v_{\rm c}$. 
 However, this scaling is only a special case.  Even with an isotropic velocity dispersion ($\beta = 0$), the spherical Jeans equation tells us that $v_c^2 \propto \Gamma  \, \sigma^2$, where $\Gamma = -\mathrm{d}\ln \rho_{\rm dm} /\mathrm{d}\ln r$ is the log-slope of the dark matter density profile.   It would unsurprising if $\Gamma$ varied systematically with $v_c$ such that $\sigma$ and $v_c$ were not linearly proportional.

Assuming Equation \ref{eqn:sigma_vs_vc} holds for the Milky Way we find  $\sigma_{\rm 3D}(v_{\rm c,MW})=\data{disp}^{+\data{ddisp_plus}}_{-\data{ddisp_minus}}\,\velunit$. This happens to be very close to $\sqrt{3/2} \,  v_{\rm c,MW}$ for our adopted Milky Way circular speed, though for much larger or smaller circular speeds, our model would give different results.

\textbf{Local escape speed:} Conventional escape-speed estimates depend on an assumption for the escape distance and have recently ranged anywhere from ${\sim}445$ to ${\sim}580\,\rm km\,s^{-1}$ \citep{piffl2014the-rave, monari2018the-escape, deason2019the-local, necib2022substructure}. Rather than choosing a distance, we measure the escape speed in solar regions in our simulations as the speed above which no particles are found. With this choice, the escape speed in our simulations is well characterized by a power-law, $v_{\rm esc} \propto v_{\rm c}^\varepsilon$, with $\varepsilon \simeq 0.5$ (Equation \ref{eqn:vesc(vc)} and Figure \ref{fig:vlim}). 
Assuming this holds for the Solar location in the Milky Way we find
 $v_{\rm esc}(v_{\rm c, MW})=\data{vesc_mw(vc)}\pm\data{dvesc_mw(vc)}\,\velunit$.

\textbf{Speed distribution:} We find that the distribution of dark matter particle speeds is well-described by a modified Maxwellian (Equation \ref{eqn:final_model}) with two shape parameters, both of which correlate with the observed $v_{\rm c}$. We use that modified Maxwellian to predict the speed distribution of dark matter near the Sun and find that it peaks at a most probable speed of $\data{v0_mw}\,\mathrm{km\,s^{-1}}$ and begins to truncate sharply above $\data{vcrit_mw}\,\mathrm{km\,s^{-1}}$. (See Figure \ref{fig:distrib_mw}.) This peak speed is somewhat higher than expected from the standard halo model, and the truncation occurs well below the formal escape speed to infinity, with fewer very-high-speed particles than is often assumed. The best fit parameters are given in Table \ref{tab:params}.

\begin{table}
\centering
\caption{Final results for the Milky Way}
\label{tab:params}
\begin{threeparttable}
\begin{tabular}{lll}
\toprule
 Parameter & Best estimate & Units \\
\midrule
$\rho(v_\mathrm{c,MW})$ & $\data{rho_GeV}\pm\data{drho_GeV}$ & $\denunit$ \\
 & $\data{rho_1e7msun}^{+\data{drho_1e7msun_plus}}_{-\data{drho_1e7msun_minus}}$ & $10^7\mathrm{M_\odot\,kpc^{-3}}$ \\
$\sigma_\mathrm{3D}(v_{\mathrm{c,MW}})$ & $\data{disp}^{+\data{ddisp_plus}}_{-\data{ddisp_minus}}$ & $\mathrm{km\,s^{-1}}$\\
$v_{\rm esc}(v_{\rm c, MW})$ &
$\data{vesc_mw(vc)}\pm\data{dvesc_mw(vc)}$ & $\velunit$\\[1.3ex]
\midrule
$v_0(v_\mathrm{c,MW})$\tnote{*} & $\data{v0_mw}\pm\data{dv0_mw}$ & $\mathrm{km\,s^{-1}}$ \\
$v_\mathrm{damp}(v_\mathrm{c,MW})$\tnote{*} & $\data{vdamp_mw}\pm\data{dvdamp_mw}$ & $\mathrm{km\,s^{-1}}$ \\[1.0ex]
\bottomrule
\end{tabular}
\begin{tablenotes}
\item[*]Shape parameters for our model of the speed distribution defined in Equation \ref{eqn:final_model}
\end{tablenotes}
\end{threeparttable}
\end{table}

This study's major contribution to the local dark matter speed distribution discussion is our methodology for predicting the Milky Way's parameter values, which slide higher or lower based on the assumed local circular speed of the Galaxy.  While we have presented a functional form that works particularly well, others could apply our methodology to other functions. In fact, we do just that for the Mao parameterization in the Appendix Section \ref{sec:Mao}.

One potential subject for further exploration is the effect of Magellanic Clouds and merger-induced streams on local dark matter speed distributions. For example, simulations have shown that both the Sagittarius dwarf \citep{purcell2012dark} and LMC \citep{besla2019the-highest-speed, donaldson2022effects, smith-orlik2023the-impact} could boost the Milky Way's high-speed tail. There is opportunity to extend an analysis like the one given here to included simulated systems with LMC-size objects and extended streams to more accurately predict the high-speed tail of dark matter particles.

\section{Acknowledgments} 
PGS and JSB were supported by NSF grant AST-1910965 and NASA grant 80NSSC22K0827. MBK acknowledges support from NSF CAREER award AST-1752913, NSF grants AST-1910346 and AST-2108962, NASA grant 80NSSC22K0827, and HST-AR-15809, HST-GO-15658, HST-GO-15901, HST-GO-15902, HST-AR-16159, HST-GO-16226, HST-GO-16686, HST-AR-17028, and HST-AR-17043 from the Space Telescope Science Institute, which is operated by AURA, Inc., under NASA contract NAS5-26555.

\appendix

\section{Fitting procedure and defining uncertainties}
\label{sec:errors}
This study develops models for predicting local dark matter density, velocity dispersion, and escape speed based on circular-speed power laws. It achieves this by performing linear regressions on the logarithms of both the target variable and circular speed. Each model has a band of uncertainty around its prediction curve. The band has two components: the standard error of the regression and the standard error of the mean. The former is
\begin{equation}
s_{\rm reg}=\sqrt{\frac{1}{N-2}\sum_i^N e_i^2}
\end{equation}
where $N$ is the number of data points, and $e_i$ is the difference between data point $i$ and the model's prediction for that point. The latter captures the increasing uncertainty in the prediction with increasing distance from the average circular speed. It is 
\begin{equation}
s_{\rm mean}(\log v_{\rm c})=\frac{s_{\rm reg}}{\sqrt{N}}\sqrt{1+\frac{(\log v_{\rm c}-\overline{\log v}_{\rm{c}})^2}{{\rm var}(
\log\boldsymbol{v_{\rm c}})}},
\end{equation}
where 
\begin{equation}
{\rm var}(\log\boldsymbol{v_{\rm c}})=\frac{\sum_i^N(\log v_{{\rm c},i}-\overline{\log v}_{\rm c})^2}{N}.
\end{equation}
The standard error of a given prediction is then $s_{\rm pred}(\log v_{\rm c})=\sqrt{s_{\rm reg}^2+s_{\rm mean}^2(\log v_{\rm c})}$. From here, the final uncertainty band around a given prediction is
\begin{align}
\delta_{\rm log\,band}(\log v_{\rm c})=&t_{\rm crit}s_{\rm pred}(\log v_{\rm c})\\
\approx&0.42\,s_{\rm pred}(\log v_{\rm c})
\end{align}
where $t_{\rm crit}$ is the critical two-tailed t-value assuming 68\% confidence and $N-2$ degrees of freedom. Keep in mind that $\delta_{\rm log\,band}(\log v_{\rm c})$ applies to the prediction line in log space. The transformation from the logarithmic error band to linear is
\begin{equation}
\delta_{\rm band}(v_{\rm c}) 
= \pm 10^{\log y \pm \delta_{\rm log\,band}(v_{\rm c})} \mp y.
\end{equation}

For our Milky Way prediction, it is also necessary to propagate the $\delta v_{\rm c,MW}=7\,\velunit$ systematic uncertainty in circular speed from ref. \cite{eilers2019the-circular} and the $\delta
\rho_{\rm data}/\rho_{\rm data}=\data{stdev_linear_dendiff}$, $\delta\sigma_{\rm data}/\sigma_{\rm data}=\data{stdev_linear_dispdiff}$ uncertainties in this study's density and dispersion data points, which stem from possible variations across the Solar ring discussed in Section \ref{sec:den_disp}. Therefore, the final uncertainty is
\begin{equation}
\delta y_{\rm MW} = \sqrt{\delta_{\rm band}^2(v_{\rm c}) + \delta v_{\rm c,MW}^2\left(\diff{y}{v_{\rm c}}\right)^2 + y^2\left(\frac{\delta y_{\rm data}}{y_{\rm data}}\right)^2}
\end{equation}
where $y$ is one of $\rho$, $\sigma_{\rm 3D}$, or $v_{\rm esc}$. Note that we do not evaluate intra-ring variations in escape speed, so we effectively assume $\delta v_{\rm esc}/v_{\rm esc}=0$.

\section{Velocity dispersion anisotropy}
\label{sec:anisotropy}

\begin{figure}
\centering
\includegraphics[width=\textwidth]{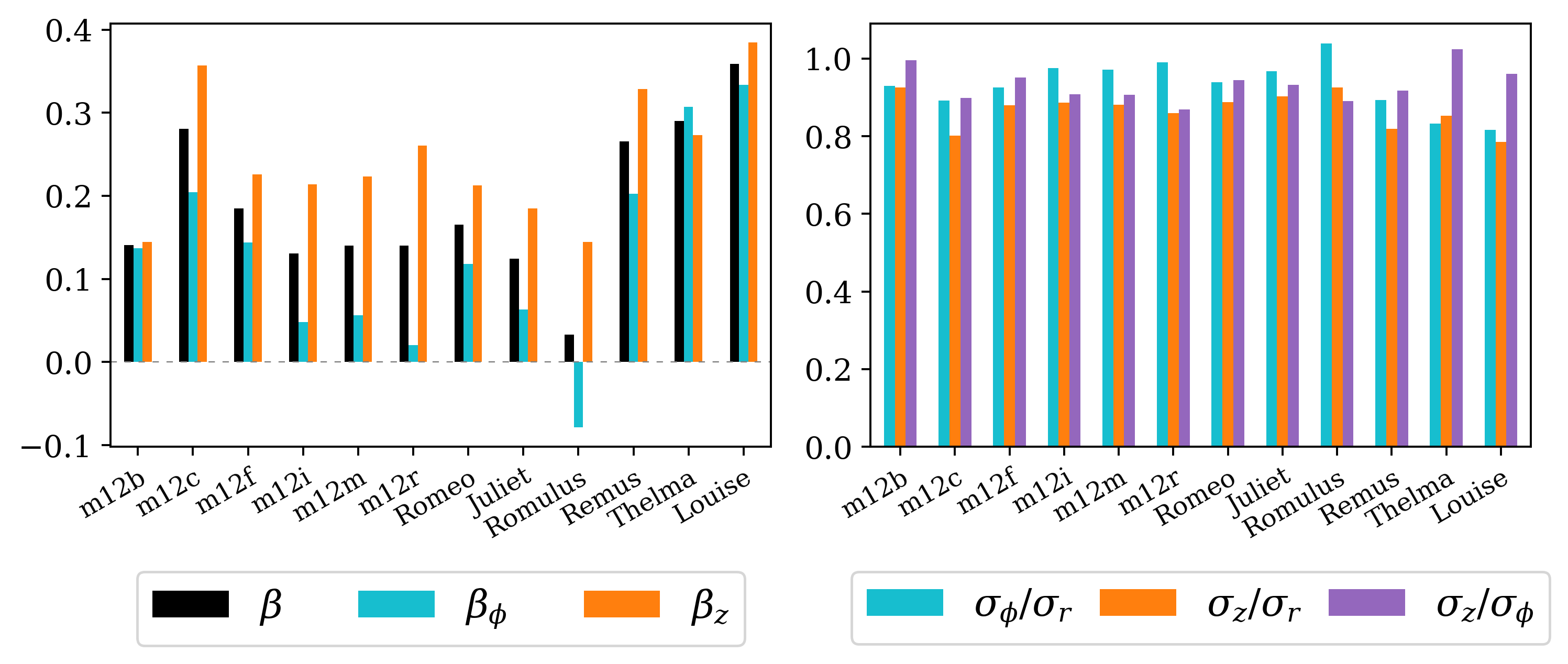}
\caption{Characterization of the anisotropy of dark matter velocity dispersion over our suite of simulations. The left panel shows the anisotropy parameters in cylindrical coordinates, defined in Equations \ref{eqn:beta}--\ref{eqn:betaz}. On the right we show the ratio of velocity dispersion components. In most cases the radial dispersion is slightly larger than the other two components, and $\sigma_{\phi}$ is larger than $\sigma_{z}$. In only one case (Thelma) is the $z$-component of dispersion larger than the $\phi$ component. Overall, anisotropy is low, with $\beta<0.2$ for the majority of galaxies in the left panel. Additionally, dispersion ratios in the right panel are all approximately between 0.8 and 1.}
\label{fig:anisotropy}
\end{figure}

Figure \ref{fig:anisotropy} summarizes the anisotropy of the local dark matter velocity dispersions in cylindrical coordinates $\phi$, $z$, and $r$. 
On the left we show the anisotropy parameters, defined as
\begin{align}
&\beta = 1 - \dfrac{\sigma_\phi^2+\sigma_z^2}{2\sigma_r^2} \label{eqn:beta}\\
&\beta_\phi = 1 - \dfrac{\sigma_\phi^2}{\sigma_r^2} \\
&\beta_z = 1 - \dfrac{\sigma_z^2}{\sigma_r^2}. \label{eqn:betaz}
\end{align}
Positive values correspond to particles moving faster radially than vertically and azimuthally, while negative values convey the opposite. Overall anisotropy is low for FIRE disks; the majority have $\beta<0.2$, and only Louise is higher than 0.3. Looking at specific components, azimuthal and radial dispersions tend to match more closely than vertical and radial, as represented by the lower cyan and higher orange bars. The right panel directly shows the ratio of the velocity dispersion components. For most galaxies, the radial dispersion is slightly larger than the other two components, and the $z$ dispersion is the smallest.  One interesting outlier is Romulus, which has a slightly larger $\phi$ velocity dispersion than radial velocity dispersion. Though some level of velocity dispersion anisotropy does exist, the ratios are all close to unity at ${\sim}0.8$--1.

\section{Using the Mao speed distribution}
\label{sec:Mao}

Note that, in the past, studies have evaluated the best-fit Mao \cite{mao2013halo-to-halo} shape parameters (Equation \ref{eqn:mao}) for individual simulated galaxies (e.g. \citep{pillepich2014the-distribution, sloane2016assessing, bozorgnia2016simulated}), but have not gone beyond implying that the best parameters for the Milky Way lie somewhere in the range of parameter values found for those simulations. In this work have aimed to provide models that can be applied to the Milky Way using the observed value of $v_c$.  In this section we do so for the Mao parameterizaton.

\begin{figure*}
\includegraphics[width=\textwidth]{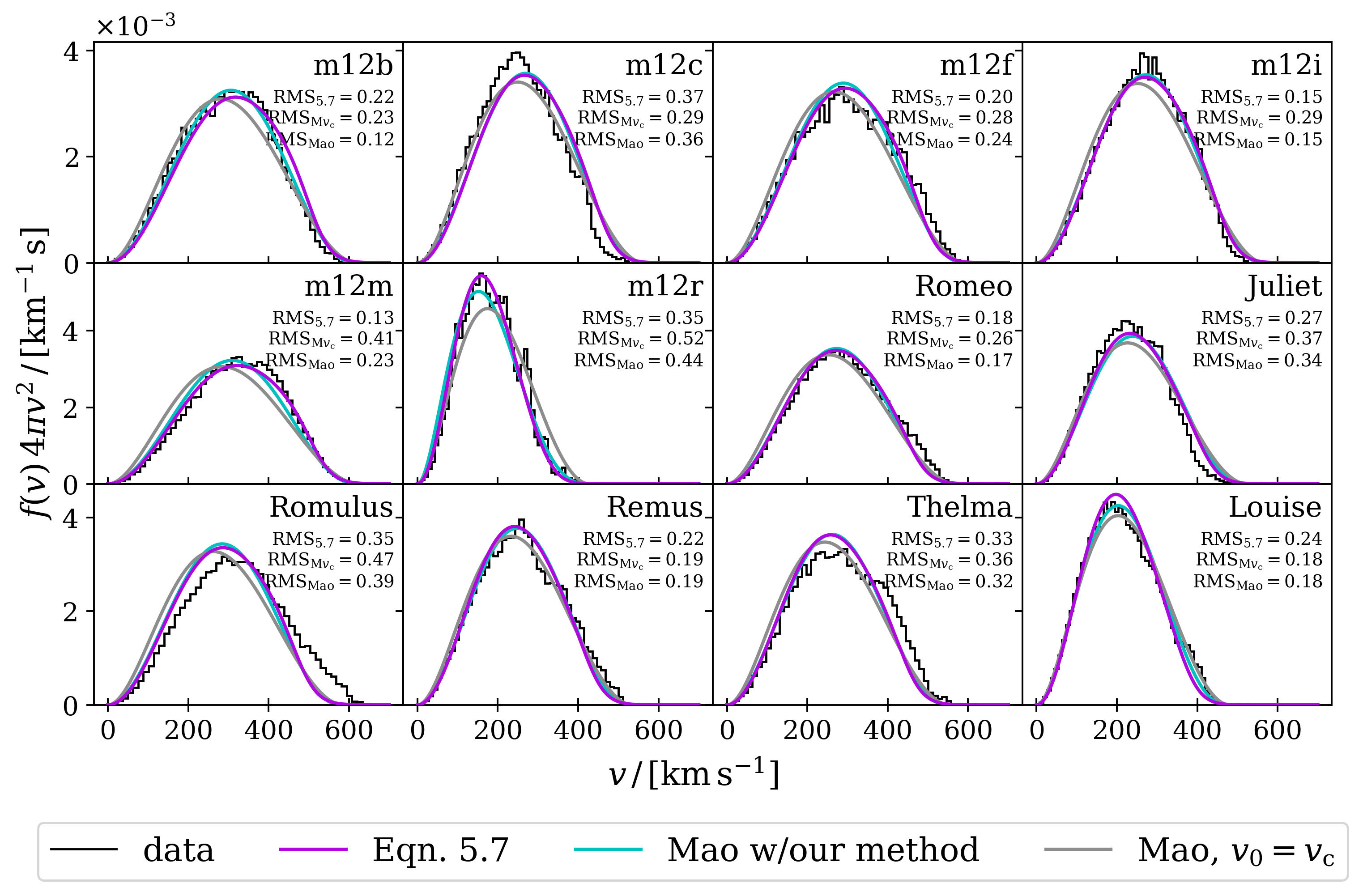}
\caption{The \staudtcolor{} curves show the final model that this study recommends, given by Equations \ref{eqn:final_model}--\ref{eqn:final_vdamp}. The grey curve shows a straightforward implementation of the Mao parameterization with the peak speed equal to the ``observed" circular speed of each galaxy. Specifically, it shows Equation \ref{eqn:mao} with parameters $(v_0, v_{\rm esc}, p)=\left(v_{\rm c},\, v_{\rm esc}(v_{\rm c}),\, \data{p_mao_naive_agg}\right)$, where Equation \ref{eqn:vesc(vc)} provides $v_{\rm esc}(v_{\rm c})$ and $p=\data{p_mao_naive_agg}$ minimizes the SSE between the model and the data for the aggregate of all twelve galaxies. The cyan curve is an implementation of the Mao parameterization that allows both $v_{\rm esc}$ and $v_0$ to vary using best-fit power laws on $v_{\rm c}$. Using Equation \ref{eqn:vesc(vc)} to determine $v_{\rm esc}$ is the same as in the standard implementation; however, we also predict $v_0$ with Equation \ref{eqn:mao_v0} as opposed to the standard $v_0=v_{\rm c}$ assumption. To be precise, our full Mao implementation uses $(v_0, v_{\rm esc}, p)=\left(v'_0(v_{\rm c}),\, v_{\rm esc}(v_{\rm c}),\, \data{p_mao_lim_fit}\right)$, where $p=\data{p_mao_lim_fit}$ minimizes the aggregate SSE. Each panel shows the root mean square errors of the three models versus the black histogram data, in units of $10^{-3}\,\rm km^{-1}\,s$. The $v_0 = v_c$ version of the Mao model performs surprisingly well, but our model offers an improvement; quantitatively, it is the most accurate of the three models shown. 
The root mean squared error for the aggregate of all twelve galaxies is $\rm{RMS}_{\ref{eqn:final_model}}=\data{staudt_rms}$ for Equation \ref{eqn:final_model}, $\rm{RMS}_{\rm Mao}=\data{mao_ours_rms}$ for Mao with our method, and $\rm{RMS}_{\rm{M}v_{\rm c}}=\data{mao_naive_rms}\ \rm km^{-1}\,s$ for Mao with $v_0=v_{\rm c}$.
}
\label{fig:mao_compare}
\end{figure*}

Figure \ref{fig:mao_compare} shows the dark matter speed distributions for each of our galaxies (black histograms) compared to best-fit models using the Mao parameterization defined by Equation \ref{eqn:mao}, along with our preferred model. We first evaluate the predictive strength of the Mao parameterization by using $v_0=v_\mathrm{c}$ and escape speed given by Equation \ref{eqn:vesc(vc)}.  For the third parameter, $p$, we make the simplifying assumption that one value should work well for all galaxies. We use MCMC to find the best-fit value of $p=\data{p_mao_naive_agg}\pm\data{dp_mao_naive_agg}$. This version of Mao performs well. Even so, our model given by Equations \ref{eqn:final_model}--\ref{eqn:final_vdamp} does a little better in most cases. It provides a ${\sim}25\%$ improvement from Mao with $v_0=v_\mathrm{c}$ and $p=\data{p_mao_naive_agg}$ judged by the root-mean-squared error of the aggregate of the predicted distributions, pushing it from $\rm{RMS_{Mao}}=\data{mao_naive_rms}\,\rm km^{-1}\,s$ down to $\rm{RMS_{5.7}}=\data{staudt_rms}\,\rm km^{-1}\,s$.

Furthermore, this study can improve the Mao implementation by predicting the best value for $v_0$ with a methodology similar to our model. In this, which we call our implementation of the Mao parameterization, 
\begin{equation}
v_0'=d'\left(\frac{v_\mathrm{c}}{100\,\mathrm{km\,s^{-1}}}\right)^{e'},
\label{eqn:mao_v0}
\end{equation}
$d'=\data{d_mao_lim_fit}^{+\data{dd_mao_lim_fit_plus}}_{-\data{dd_mao_lim_fit_minus}}\,\velunit$, and $e'=\data{e_mao_lim_fit}\pm\data{de_mao_lim_fit}$. Additionally, the best fit value for $p$ is
\begin{equation}
p=\data{p_mao_lim_fit}\pm\data{dp_mao_lim_fit}.
\label{eqn:p_ours}
\end{equation}
This framework performs very well, as shown by the cyan line in Figure \ref{fig:mao_compare}, although our preferred model tends to perform slightly better; the root mean square error for our implementation of Mao across the aggregate of all twelve galaxies is $\mathrm{RMS}_\mathrm{Mao}=\data{mao_ours_rms}\,\mathrm{km^{-1}\,s}$. Although we do prefer the model described by Equations \ref{eqn:final_model}--\ref{eqn:final_vdamp} given its higher performance, the results here show that if one's preference is to use Mao, one could achieve almost as accurate results by implementing our methods of determining $v_{\rm esc}(v_{\rm c})$ with Equation \ref{eqn:vesc(vc)} and predicting $v_0$ for Mao using Equation \ref{eqn:mao_v0}.

The MCMC methodology for these two implementations of the Mao model is similar to that which this study uses  to fit our preferred model, described in Section \ref{sec:mcmc}. For the Mao optimizations, we substitute Equation \ref{eqn:mao} for $f$ and use the appropriate parameter vector $\theta$, assuming flat priors. For the standard $v_0=v_{\rm c}$ Mao model, we use $p\in[1, 5]$. For our implementation of Mao, $d\in[10, 130]\,\rm km\,s^{-1}$, $e\in[0.8, 5]$, and $p\in[1, 5]$.

\section{Circular Speed Metric}

\begin{figure}
\includegraphics[width=\textwidth]{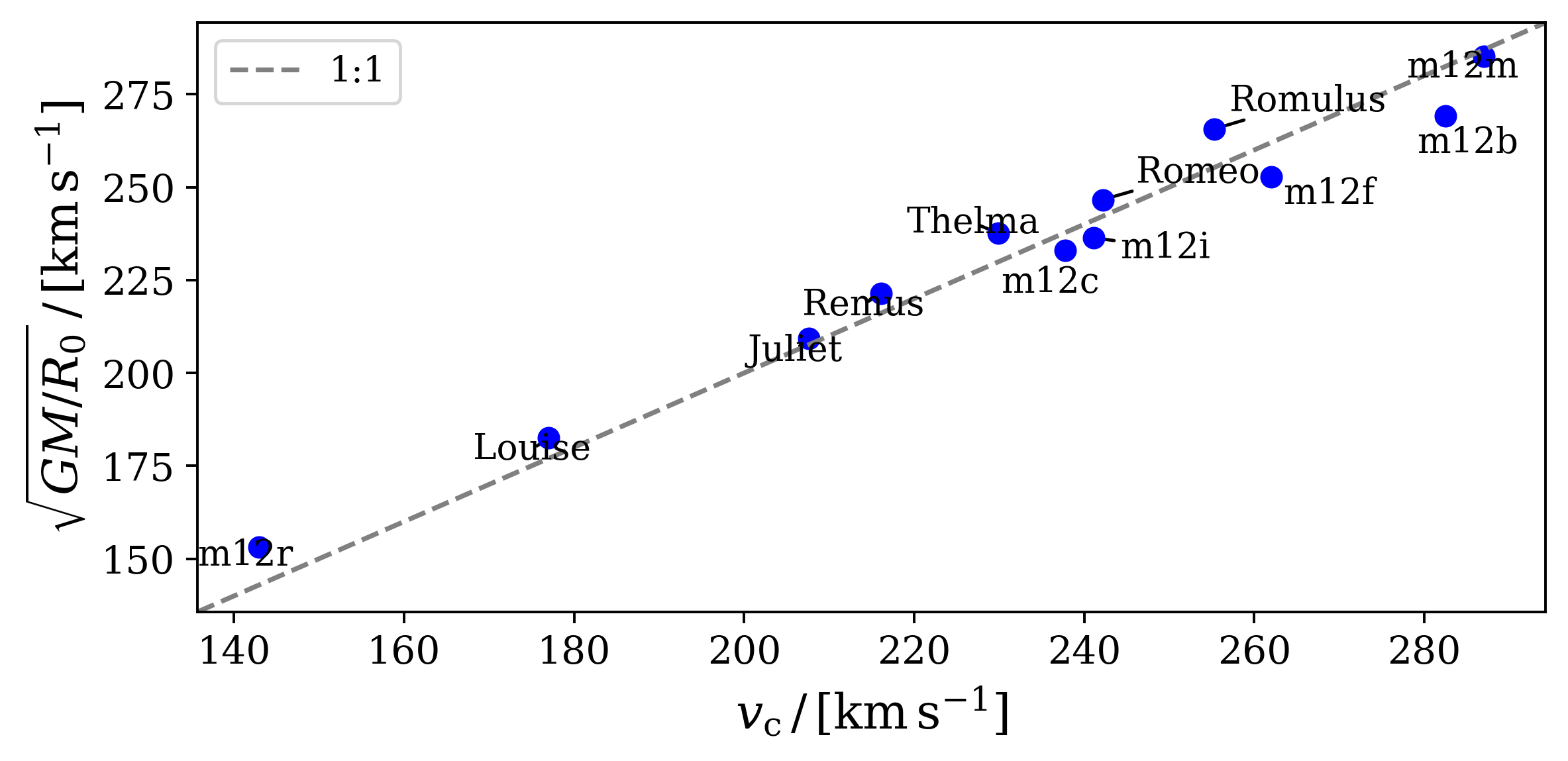}
\centering
\caption{Here we compare measured values of $v_\mathrm{c}$ (the average $\phi$ velocity of cold gas $v_\mathrm{c}=\langle\vec{v}\cdot\hat{\phi}\rangle_{T\leq 10^4\mathrm{K}}$) in our simulations to a spherical idealization of the circular speed $\sqrt{GM/R_0}$. The one-to-one line on this scatter plot has a 0.96 $r^2$ coefficient of determination. This suggests that the detailed choice of how we measure or characterize $v_\mathrm{c}$ in the simulations is probably not crucial to the success of our power-law characterizations in the paper.
}
\label{fig:gmr_vs_vc}
\end{figure}

For completeness we provide Figure \ref{fig:gmr_vs_vc}, which plots this study's observable proxy for circular speed, $v_\mathrm{c}=\langle\vec{v}\cdot\hat{\phi}\rangle_{T\leq 10^4\mathrm{K}}$, versus the spherical ideal $\sqrt{GM/R_0}$. 
The two exhibit a near one-to-one relationship, with only a 4\% scatter. Therefore, this study's results are likely insensitive to the choice of whether to use the ideal or observable circular speed.

\section{Halo Integrals}

\begin{figure*}
\includegraphics[width=\textwidth]{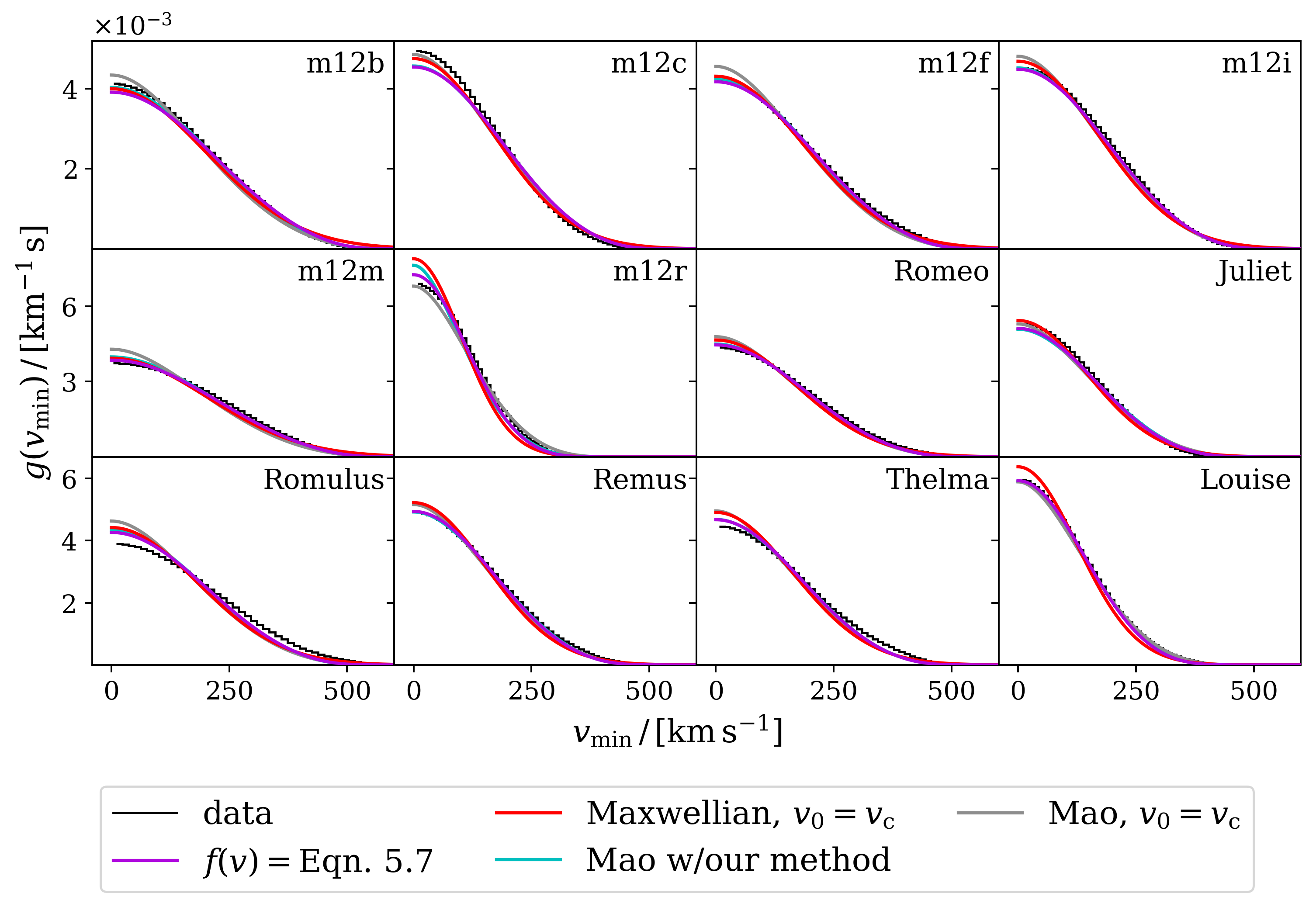}
\caption{The \staudtcolor{} lines represent the halo integrals for this work's speed-distribution model given by Equations \ref{eqn:final_model} through \ref{eqn:final_vdamp}. The \maxwellcolor{} lines represent that given by the simple Maxwellian. The grey lines represent a standard implementation of the Mao model, where $v_0=v_{\rm c}$, and we use our method of predicting the escape speed with Equation \ref{eqn:vesc(vc)}. The truncation-strength $p$ takes the best fit value of \data{p_mao_naive_agg}. The cyan lines show Mao but using this study's method of predicting the peak speed with Equation \ref{eqn:mao_v0} and predicting the escape speed with Equation \ref{eqn:vesc(vc)}. $p$ takes the best-fit value of \data{p_mao_lim_fit}.
}
\label{fig:halo_integrals_all}
\end{figure*}

Figure \ref{fig:halo_integrals_all} compares halo integral performance for our model in \staudtcolor{} verses a Maxwellian in \maxwellcolor{}. The Maxwellian's halo integral tends to exhibit an excess at low speeds while falling too low in the intermediate range. Our model's integral provides better agreement with the data.

\nocite{sofue2020rotation, necib2019under}

\bibliographystyle{JHEP}
\bibliography{bib.bib}

\end{document}